\documentclass[reprint,showpacs,
amsmath,amssymb,aps,pra]{revtex4-1}

\usepackage{dcolumn}
\usepackage{bm}

\usepackage{amsfonts}
\usepackage[mathscr]{euscript}
\usepackage{mathrsfs}

\newcommand{\Tr}{\operatorname{Tr}}
\newcommand{\rmd}{\mathrm{d}}
\newcommand{\rme}{\mathrm{e}}
\newcommand{\rmi}{\mathrm{i}}
\newcommand{\Acal}{\mathcal{A}}

 \newcommand{\Ecal}{\mathcal{E}}

\newcommand{\Ical}{\mathcal{I}}

\newcommand{\Lcal}{\mathcal{L}}

\newcommand{\Scal}{\mathcal{S}}

 \newcommand{\Dscr}{\mathscr{D}}
 
\newcommand{\Fscr}{\mathscr{F}}
\newcommand{\Gscr}{\mathscr{G}}
\newcommand{\Hscr}{\mathscr{H}}

\newcommand{\Lscr}{\mathscr{L}}

\newcommand{\Sscr}{\mathscr{S}}
\newcommand{\Tscr}{\mathscr{T}}

\newcommand{\Ebb}{\mathbb{E}}
\newcommand{\Var}{\operatorname{Var}}
\newcommand{\Cov}{\operatorname{Cov}}

\newcommand{\Rbb}{\mathbb{R}}
\newcommand{\Cbb}{\mathbb{C}}

\newcommand{\Qbb}{\mathbb{Q}}
\newcommand{\Pbb}{\mathbb{P}}
\newcommand{\norm}[1]{\left\Vert#1\right\Vert}
\newcommand{\abs}[1]{\left\vert#1\right\vert}

\newcommand{\RE}{\operatorname{Re}}
\newcommand{\IM}{\operatorname{Im}}

\begin{document}
\title{Quantum trajectories: memory and continuous observation}

\author{Alberto Barchielli}
\affiliation{Politecnico di Milano, Dipartimento di Matematica, Piazza Leonardo da
Vinci 32, I-20133 Milano, Italy} \altaffiliation[Also at ]{Istituto Nazionale di
Fisica Nucleare (INFN), Sezione di Milano, and Istituto Nazionale di Alta Matematica
(INDAM-GNAMPA).}
\author{Cl\'ement Pellegrini}
\affiliation{Laboratoire de Statistique et Probabilit\'es, Universit\'e Paul Sabatier,
118, Route de Narbonne, 31062 Toulouse Cedex 4, France.}
\author{Francesco Petruccione}
\affiliation{University of KwaZulu-Natal,
School of Physics and National Institute for Theoretical Physics, Private Bag X54001, Durban 4000, South Africa.}
\date{\today}

\begin{abstract} Starting from a generalization of the quantum trajectory theory [based on the stochastic Schr\"odinger equation (SSE)], non-Markovian models of quantum dynamics are derived. In order to describe non-Markovian effects, the approach used in this article is based on the introduction of random coefficients in the usual linear SSE. A major interest is that this allows a consistent theory of quantum measurement in continuous time to be developed for these non-Markovian quantum trajectory models. In this context, the notions of `instrument', `\emph{a priori}', and `\emph{a posteriori}' states can be introduced. The key point is that by starting from a stochastic equation on the Hilbert space of the system, we are able to respect the complete positivity of the mean dynamics for the statistical operator and the requirements of the axioms of  quantum measurement theory.
The flexibility of the theory is next illustrated by a concrete physical model of a noisy oscillator where non-Markovian effects come from the random environment, colored noises, randomness in the stimulating light, and delay effects. The statistics of the emitted photons and the heterodyne and homodyne spectra are studied, and we show how these quantities are sensitive to the non-Markovian features of the system dynamics, so that, in principle, the observation and analysis of the fluorescent light could reveal the presence of non-Markovian effects and allow for a measure of the spectra of the noises affecting the system dynamics.
\end{abstract}

\pacs{42.50.Lc, 03.65.Ta, 03.65.Yz }

\maketitle
\section{Introduction}

A first aim of the theory of open quantum systems is the description of the time
evolution of a  quantum system $\Scal$ (the open system) interacting with an
environment $\Ecal$ \cite{BreP02}. More precisely, one focuses on the reduced
evolution of $\Scal$ after tracing out the degrees of freedom of $\Ecal$. The
resulting evolution is then usually described in terms of generalized \emph{master equations}
for the reduced density matrix $\rho(t)$.

A particular simple and useful way to describe an open system is provided by the
Markovian approximation \cite{Carm02}. Essentially, this approach is based on the absence of memory
effects in the environment. In this situation, the master equations are
linear first-order differential equations with a possibly time dependent generator.  The
generator takes a particular form, the well known \emph{Lindblad} form, that
guarantees the \emph{complete positivity} of the dynamics, as required by quantum
mechanics.

Unfortunately, this approximation is no longer valid when the memory effects of the
environment cannot be neglected. Several physical situations involve long-memory-time effects and lead to a non-Markovian behavior: strong coupling,
correlation, and entanglement in the initial $\Scal$-$\Ecal$ state \cite{SBPV}, a? system at low
temperature \cite{Weiss}, and a? structured environment \cite{Lambro}. In this context, the master equations take
different forms according to the physical situation, e.g., integro-differential equations
\cite{BreP02, Smirne}, time-convolutionless equations \cite{BreP02}, and Lindblad rate
equations \cite{Bu1}. The common point is that they are not in the
Lindblad form, which is characteristic of Markovian evolutions.

In both situations (Markovian and non-Markovian) a useful approach  to
describe concrete physical evolutions is provided by the theory of the \emph{stochastic
Schr\"odinger equation} (SSE) \cite{Bel88,Dio88a,BarB91,Carm}.  A SSE is a nonlinear stochastic
differential equation for a wave-function process $\psi(t)$. The link with the
traditional master equation is given by the average property
$\Ebb[\vert\psi(t)\rangle\langle\psi(t)\vert]=\rho(t)$, where $\Ebb$ denotes the
average over the realizations of $\psi(t)$. To find the SSE providing a given master
equation by averaging is called \emph{unraveling}. The idea of unraveling has been
a real breakthrough for simulating master equations; it is at the root of the
\emph{Monte-Carlo wave function method} \cite{BreP02,MCwfm}. Indeed, for huge systems, the
description of $\psi(t)$ requires many fewer parameters than the ones needed for
$\rho(t)$.

However, the construction of adequate SSEs has been essential also for a second
aspect of the theory of open quantum systems: the description of the monitoring of
$\Scal$. In special situations, the SSE can be interpreted in terms of quantum
measurements. More precisely, in these cases, the solution $\psi(t)$ is called a
\emph{quantum trajectory} and describes the evolution of an open system undergoing
indirect continuous measurement (continuous monitoring)
\cite{GarZ04,WisM10,BarG09,BarB91}. In particular the noises involved in the SSE,
describing jump or diffusion evolutions, can be directly connected with the outputs
of measurement apparatuses. Such an interpretation is crucial in the understanding
of real quantum optics experiments \cite{Carm08,BarG09,BarB91,Bar90,Carm} such as direct
photo-detection, spectral photo-detection, homodyning, and heterodyning. They are also at the cornerstone of
modern technology such as feedback control \cite{WisM10,Bel88,WWM}. As a
consequence an active line of research consists in finding SSEs that can be
physically interpreted in terms of continuous monitoring of the system.

In the Markovian case, this link is clearly established for almost all situations
\cite{GarZ04,WisM10,BarG09}.
Starting from a master equation in Lindblad form, it is known how to construct an
appropriate unraveling in terms of a SSE. The (nonlinear) SSE is a stochastic
equation for a random normalized vector $\psi(t)$. It is always possible to
construct a linear SSE, driven by Poisson and Wiener noises, for a non-normalized
vector $\phi(t)$, such that $\psi(t)=\phi(t)/\norm{\phi(t)}$. Moreover, the linear
and nonlinear versions of the SSE are related by a change of probability measure and
it is this link that allows for a measurement interpretation consistent with the
postulates of quantum mechanics \cite{BarG09,BarB91}. In mathematical terms the
change of measure is a \emph{Girsanov transformation} with probability density
$\norm{\phi(t)}^2$; the key point that allows this transformation is the fact that
$\norm{\phi(t)}^2$ turns out to be a \emph{martingale} \cite{BarG09}. Moreover,
these stochastic differential equations can be deduced from purely quantum evolution
equations for the measured system coupled with a quantum environment, combined with a
continuous monitoring of the environment itself \cite{quantum,Castro}.

In the non-Markovian case, to find relevant SSEs, describing both non-Markovian
quantum evolutions and continuous monitoring, is a tremendous challenge. In contrast to
the Markovian framework, no general theory has been developed. Essentially there exist two
strategies.

The first strategy consists in considering a physical model described by a
non-Markovian master equation and in finding an appropriate pure-state unraveling.
This approach has been successfully applied in various situations. A major common
point consists in replacing the memoryless white noises, used in Markovian SSEs, by
colored Gaussian noises and in introducing some delay effects \cite{Stru1}. This allows the
introduction of correlations in time that describe strong memory effects of the
environment interaction. In this direction several models, such as, for example, the so-called \textit{non-Markovian quantum state diffusion}, have been derived \cite{Stru3}.
Other investigations involving non-Markovian jump type SSEs have been also proposed
\cite{Piilo1,Piilo3}. While such approaches are efficient for simulating relevant non-Markovian
evolutions, the measurement interpretation of the underlying SSEs is highly debated
\cite{Gam4,Diosi3,Piilo3} and a complete conclusion is still lacking. A principal problem concerns
the interpretation of the underlying noises as outputs of continuous-time measurements. For other models
such as Lindblad rate equations, different types of jump unravelings have been proposed \cite{MoPe}.
Jump-diffusion generalizations with measurement applications have been derived in Ref.\ \cite{BarPel}.
In this context limitations also appear in the sense that the types of observables that can be measured must
have particular and restrictive forms.

A second strategy is first to generalize directly the Markovian SSE by introducing
memory effects. Then, one has to show whether this SSE provides the unraveling of
some non-Markovian evolution and whether it has a physical
measurement interpretation. To work at the Hilbert space level guarantees
automatically the complete positivity of the evolution of the statistical operator.
In this paper we propose non-Markovian SSE models with physical measurement
interpretations. Our strategy consists in adapting the Markovian approach by
replacing white and Poisson noises with non-Markovian noises and by allowing for
random coefficients in the equation. First we start with a linear SSE driven by
colored noises and involving random operator coefficients; the whole randomness is
defined under a reference probability. Next we introduce the physical probability and
the nonlinear SSE, which is a stochastic differential equation under the new probability.
Finally it is possible to pass to the linear and nonlinear versions of the stochastic master equation
(SME), and we show that they determine the dynamics and the continuous measurements without violating the
axiomatic structure of quantum mechanics. The general mathematical structure was introduced in \cite{BarH95,BDPP11};
in \cite{BPP10} we started to show how such a structure allows for the introduction of some colored noises, while in
\cite{BarG11} we considered memory effects due to feedback with delay. The present article is devoted to the
exploration and clarification of  physical effects that can be treated within such a theory and to
show them on a concrete physical system.

The paper is structured as follows. Section \ref{sec:SSE} describes the general theory of the stochastic Schr\"odinger
equation. We present the general mathematical ingredients necessary to develop
the generalization of the SSE involving colored noises and random coefficients. We consider first a linear
stochastic equation for a non normalized wave function $\phi(t)$. Then, with the
help of a change of measure, determined by the SSE itself, we derive the nonlinear SSE for
the wave function $\psi(t)=\phi(t)/\norm{\phi(t)}$. In Sec.\ \ref{sec:sme} the linear SME and the nonlinear one are
introduced and the measurement interpretation is justified by introducing positive operator-valued measures,
instruments, \emph{a priori} states (mean states), and \emph{a posteriori} states (conditional states). Section \ref{sec:model} is devoted to a
concrete model, a noisy oscillator absorbing and emitting light, by which physical effects can be discussed and the possibilities of the theory can be explored. Moreover, in this section we study the
behavior of the outputs of the oscillator; in particular we study the effects of the non-Markovian terms in the
dynamics on the homodyne and the heterodyne spectra of the emitted light and on the statistics of the photons, now
analyzed by direct detection. Conclusions are presented in Sec.\ \ref{sec:concl}.

\section{The stochastic Schr\"odinger equation}\label{sec:SSE}

When introducing non-Markovian evolutions for a quantum state, the first
problem is to guarantee the complete positivity of the evolution of the state
(statistical operator) of the reduced system. Then, if one wants to introduce
measurements in continuous time, the second problem is to have equations compatible
with quantum measurement theory. Starting from the linear version of the SSE allows
memory to be introduced by using random coefficients and colored noises (no problem of complete positivity
because we are working at Hilbert space level) and the
\emph{instruments} related to the continuous monitoring to be constructed (no problem with the
axioms of quantum theory because we are respecting linearity) \cite{BarH95,
BPP10, BDPP11}.

Let  us denote by $\Hscr$ the Hilbert space of the quantum
system of interest, a separable complex Hilbert space, by $\Lscr(\Hscr)$ the space of the bounded
operators on $\Hscr$, by $\Tscr(\Hscr)\subset \Lscr(\Hscr)$ the trace class and by
$\Sscr(\Hscr)\subset \Tscr(\Hscr)$ the convex set of the statistical operators.

\subsection{The linear SSE and the reference probability \label{sec:SSEa}}

The starting point of the whole construction is the linear SSE for a stochastic process
$\phi(t)$ with values in $\Hscr$:
\begin{multline}\label{lSSE1}
\rmd \phi(t)= K_0(t)\phi(t_-) \rmd t + \sum_{j=1}^m L_j^0(t)\phi(t_-)\, \rmd M_j(t)
\\{}+ \sum_{k=1}^{d'} \bigl(R_k(t)-\openone\bigr)\phi(t_-)\,\rmd N_k(t).
\end{multline}

In the Markovian case the $M_j$ are Wiener processes, the $N_k$ are Poisson
processes, and all these processes are independent. Moreover, the operators $R_k(t)$,
$L_j^0(t)$, and $K(t)$ are not random. The key property that allows for a
measurement interpretation is that $\norm{\phi(t)}^2$ is a mean-1
\emph{martingale}, and this requirement imposes a link among the operators $R_k(t)$,
$L_j^0(t)$, and $K_0(t)$. The non-Markovian generalization is to take more general
processes as driving noises and to allow for random coefficients. Now we illustrate
the precise meaning of the various quantities appearing in the SSE \eqref{lSSE1}.

First of all we work in a reference probability space $(\Omega, \Fscr, \Qbb)$;
$\Omega$ is the sample space, $\Fscr$ the $\sigma$-algebra of events, and $\Qbb$ a reference probability. The physical
probability will appear when the measuring interpretation is constructed in
Sec.\ \ref{sec:SSEb}. Past and present up to time $t$ are represented by the
events in the sub-$\sigma$-algebra $\Fscr_t\subset \Fscr$; the family
$(\Fscr_t)_{t\geq 0}$ is a filtration of $\sigma$-algebras satisfying the usual
hypotheses, i.e., $\Fscr_s\subset \Fscr_t$ for $0\leq s <t$, $A\in\Fscr$ with
$\Qbb(A)=0$ implies $A\in \Fscr_0$, and $\Fscr_t=\bigcap_{T>t}\Fscr_T$. In $(\Omega,
\Fscr, (\Fscr_t)_{t}, \Qbb)$ we have $d$ continuous, independent, adapted, standard
Wiener processes $B_1, \ldots, B_d$ and $d'$ adapted c\`{a}dl\`{a}g counting process of
stochastic intensities $i_k(t)\geq 0$, that are c\`{a}gl\`{a}d. The French acronym c\`{a}dl\`{a}g means with trajectories
continuous from the right and with limits from the left, while c\`{a}gl\`{a}d means continuous from the left and with
limits from the right. The meaning of stochastic intensity is given
by the heuristic conditional expectation
\begin{equation}\label{stochint}
\Ebb_\Qbb[\rmd N_k(t)|\Fscr_t]=i_k(t)\rmd t;
\end{equation}
the stochastic intensities determine the probability law of the counting processes \cite{DalVJ03}.
Assuming the usual hypothesis, that the processes are c\`{a}dl\`{a}g or c\`{a}gl\`{a}d, etc., are mathematical
regularity requirements useful in a rigorous development of stochastic calculus,
what is physically important is to have non anticipating (= adapted) processes.

The $m$ continuous processes $M_j(t)$ are given by
\begin{equation}\label{eq:M}
M_j(t) =  \int_0^tf_j(s) \rmd s + \sum_{i=1}^d \int_0^tb_{ji}(s)\rmd B_i(s),
\end{equation}
where $f_j(t)$ and $b_{jk}(t)$ are complex, adapted c\`{a}gl\`{a}d processes such that $\forall t>0$,
with probability 1, $\int_0^t \abs{f_j(t)}\rmd t<+\infty$ and $\int_0^t
\abs{b_{ji}(t)}^2\rmd t<+\infty$. Some typical choices are given in Sec.\ \ref{sec:model}.

The functions $t\mapsto L_j^0(t) $, and $t\mapsto R_k(t)$, $t\mapsto K_0(t) $ are strongly c\`{a}gl\`{a}d,
bounded operator-valued adapted processes; to be bounded is a sufficient condition
to have a well-defined general equation \cite{BarH95}. For physical problems
also the unbounded case is important and, indeed, the examples we shall give involve
unbounded operators; the case involving unbounded operators, but restricted to a Markovian dynamics, is treated in \cite{Hol1,Castro}. By allowing for random system operators and the general noises
\eqref{eq:M}, it is possible to describe random external forces, random
environments, colored baths, stochastic control, adaptive measurements and so on.

The SSE \eqref{lSSE1} is a linear stochastic differential equation in the It\^o sense. The
initial condition is taken to be
\begin{equation}\label{in_cond}
\phi(0)=\phi_0 \qquad \text{with} \qquad\Ebb_\Qbb[\norm{\phi_0}^2]=1.
\end{equation}
Note that, by suitably choosing $\Fscr_0$, $\Qbb$ and $\phi_0$, any statistical
operator $\rho_0\in \Sscr(\Hscr)$ can be represented as
$\rho_0=\Ebb_\Qbb[|\phi_0\rangle\langle \phi_0|]$. The solution $\phi(t)$ is taken
to be c\`{a}dl\`{a}g and it is unique \cite{BarH95}. To write $\phi(t_-)$ means to take the
value of $\phi$ just before the possible jump at time $t$ due to the counting
processes.

By using the explicit expressions for the processes $M_j$, the linear SSE can be
rewritten as
\begin{multline}\label{lSSE2}
\rmd \phi(t)= K(t)\phi(t_-) \rmd t + \sum_{i=1}^d L_i(t)\phi(t_-)\, \rmd B_i(t)
\\ {}+ \sum_{k=1}^{d'} \bigl[R_k(t)-\openone\bigr]\phi(t_-)\,\rmd N_k(t),
\end{multline}
where $\displaystyle
K(t)=K_0(t)+\sum_{j=1}^m f_{j}(t)L_j^0(t)$ and
\begin{equation}\label{L_i}
L_i(t)=\sum_{j=1}^m L_j^0(t)b_{ji}(t).
\end{equation}

For the physical interpretation in terms of measurements, we need
$\norm{\phi(t)}^2$ to be a martingale:
$\Ebb_\Qbb\left[\norm{\phi(t)}^2\big|\Fscr_s\right]= \norm{\phi(s)}^2$, for $0\leq s
<t$  \cite{BarG09}. We shall see that this is crucial for defining physical
probabilities. To this end we have to compute $\rmd\norm{\phi(t)}^2$. Here and in
all the formulas involving stochastic differentials, we have to use It\^o's formula
and the rules of stochastic calculus, which are summarized by It\^o's table
\[
\rmd B_i(t)\, \rmd B_j(t)=\delta_{ij}\,\rmd
t, \qquad
\rmd N_k(t)\rmd N_j(t)=\delta_{kj}\,\rmd t,
\]
\[
\rmd B_i(t)\,\rmd N_k(t)=\rmd B_i(t)\,\rmd t=\rmd N_k(t)\,\rmd t=0.
\]
Then, we get
\begin{multline}\label{dnorm2}
\rmd \norm{\phi(t)}^2= \sum_{i=1}^d\langle \phi(t_-) |
\big[L_i(t)+L_i(t)^\dagger\big] \phi(t_-) \rangle \rmd B_i(t)
\\ {}+\big\langle \phi(t_-)\big| \Bigl(K(t)^\dagger + K(t) +\sum_{i=1}^d L_i(t)^\dagger
L_i(t)\Bigr)\phi(t_-)\big\rangle \rmd t
\\ {}+\sum_{k=1}^{d'}
\bigl(\norm{ R_k(t) \phi(t_-)}^2 -\norm{\phi(t_-)}^2\bigr)  \rmd N_k(t).
\end{multline}
The martingale property is ensured if we have $\Ebb_\Qbb\left[\rmd
\norm{\phi(t)}^2\big|\Fscr_t\right]=0$. By \eqref{stochint} and $\Ebb_\Qbb[\rmd
B_i(t)|\Fscr_t]=0$, we get the restriction
\begin{multline}\label{K}
K(t)=-\rmi H(t) -\frac{1}2\sum_{i=1}^d L_i(t)^\dagger L_i(t)\\ {}+\frac 1
2 \sum_{k=1}^{d'} i_k(t)\left(\openone- R_k(t)^\dagger R_k(t)\right)
\end{multline}
with $ H(t)^\dagger=H(t)$.

\subsection{The nonlinear SSE and the physical probability \label{sec:SSEb}}
Let us define the quantity
\begin{equation}\label{newdensity}
p(t):=\norm{\phi(t)}^2, \qquad \forall t\geq 0,
\end{equation}
and the normalized version of $\phi(t)$,
\begin{equation}\label{def:psi}
\psi(t,\omega):=\begin{cases} \norm{\phi(t,\omega)}^{-1} \phi(t,\omega) & \text{if } p(t,\omega)\neq 0,
\\
z & \text{if } p(t,\omega)= 0 ,\end{cases}
\end{equation}
where $z\in \Hscr$ is a non random vector with $\norm z=1$ and we denote by $\omega$
the generic sample point in $\Omega$, as usual. Moreover, we introduce the processes
\begin{equation}\label{mi}
m_i(t):=2\RE\langle \psi(t_-)|L_i(t)\psi(t_-)\rangle ,
\end{equation}
\begin{equation}\label{jk}
j_k(t):=i_k(t)\norm{R_k(t)\psi(t_-)}^2.
\end{equation}

By condition \eqref{K}, Eq.\ \eqref{dnorm2} becomes
\begin{multline}\label{dp(t)}
\rmd p(t)= p(t_-)\biggl\{\sum_{i=1}^d m_i(t) \rmd B_i(t)\\
{}+ \sum_{k=1}^{d'}\left( \norm{R_k(t)\psi(t_-)}^2-
1\right)\bigl(\rmd N_k (t)-i_k(t)\rmd t\bigr)\biggr\}.
\end{multline}
As already said, the key property of quantum trajectory theory is that $p(t)$ is a
mean-1 $\Qbb$ martingale, which follows from this equation and the normalization
\eqref{in_cond} of the initial condition \cite[Theorem 2.4, Sec.\ 3.1]{BarH95}.

\paragraph{The physical probability.} Now we introduce the new probability measures,
whose physical meaning will be discussed in Sec.\ \ref{sec:sme}: $\forall A\in \Fscr_T$,
\begin{equation}\label{newprob}
\Pbb^T_{\phi_0}(A):=\Ebb_\Qbb[p(T)1_A]= \int_A p(T;\omega) \Qbb(\rmd \omega).
\end{equation}
Owing to the martingale property of the probability density $p(t)$, the probabilities
$\Pbb^T_{\phi_0}$ are consistent, in the sense that
$\Pbb^t_{\phi_0}(F)=\Pbb^s_{\phi_0}(F)$ for $F\in \Fscr_s$, $t\geq s \geq 0$.

The new probability $\Pbb^T_{\phi_0}$ modifies the distribution of the processes
$B_i$ and $N_k$. A very important property is that a Girsanov-type theorem holds
\cite[Proposition 2.5, Remarks 2.6 and 3.5]{BarH95}.

\paragraph{Girsanov transformation.}\label{par:Gir} Under $\Pbb^T_{\phi_0}$, in the time interval
$[0,T]$, the processes
\begin{equation}\label{newW}
W_j(t):= B_j(t)-\int_0^t m_j(s)\rmd   s, \qquad j=1,\ldots, d,
\end{equation}
are independent Wiener processes, while the counting processes $N_1,\ldots,N_{d'}$ change their
stochastic intensities, which become $j_1,\ldots,j_{d'}$. The quantities $m_i$ and $j_k$ are
defined in Eqs.\ \eqref{mi} and \eqref{jk}.

Note that, if for a certain index $i$ we have $m_i(t)=0$, $\forall t\geq 0$, then
$W_i(t)=B_i(t)$: the process $B_i$ remains a Wiener process also after the change of
probability and it is independent from all the other components of $W$. For instance, from Eq.\
\eqref{mi} we have $m_i\equiv 0$ for all initial conditions when the operator $\rmi
L_i(t)$ is self-adjoint for all $t\geq 0$.

\paragraph{The nonlinear SSE.} Under $\Pbb^T_{\phi_0}$,
in the time interval $[0,T]$, the random normalized vector \eqref{def:psi} satisfies the stochastic
differential equation
\begin{multline}\label{nlSSE}
\rmd \psi(t) = \hat K(t)\psi(t_-) \rmd t \\ {}+ \sum_{i=1}^d \left( L_{i}(t) -\frac
1 2 \, m_i(t)\right)\psi(t_-) \rmd W_{i}(t) \\ {}+\sum_{k=1}^{d'} \left(\frac
{R_k(t)\psi(t_-)} {\norm{R_k(t)\psi(t_-)}} -\psi(t_-)\right)\rmd N_k(t),
\end{multline}
with $\psi(0)=\phi_0$, and
\begin{multline}
\hat K(t):=-\rmi H(t) - \frac 1 2 \sum_{i=1}^d \left(L_i(t)^\dagger-m_i(t)\right) L_i(t) \\ {}-
\sum_{i=1}^d\frac{ m_i(t)^2}8+ \frac
12\sum_{k=1}^{d'} \bigl(j_k(t)-i_k(t) R_k(t)^\dagger R_k(t)\bigr).
\end{multline}
To get this result one needs to compute $\rmd \left(1/\sqrt{p(t)}\right)$ from Eq.\
\eqref{dp(t)} and to express this differential and $\rmd \phi(t)$ in terms of the
new Wiener processes; the rigorous proof is given in Ref.\ \cite{BarH95}.

At least in the Markov case, it is this equation that is the starting point for
powerful numerical methods \cite{BreP02, MCwfm}.

\section{The stochastic master equation}\label{sec:sme}
Now that we have presented the theory of the stochastic Schr\"odinger equation for
pure states, we develop the analog for density matrices and we introduce the
\textit{stochastic master equation}.

\subsection{The linear SME}\label{sec:lSME}
As in the case of the SSE, we start with a linear equation.
More precisely, from Eqs.\ \eqref{lSSE2} and \eqref{K} we can derive the linear SME for the process
$\tilde \sigma(t):=|\phi(t)\rangle \langle
\phi(t)|$, $t\geq0$:
\begin{multline}\label{lSME}
\rmd \tilde \sigma(t)=\Lcal(t)[\tilde \sigma(t_-)]\rmd t+ \sum_{i=1}^d \bigl( L_i(t)
\tilde \sigma(t_-) \\ {}+ \tilde \sigma(t_-) L_i(t)^\dagger\bigr)\rmd
B_i(t)+\sum_{k=1}^{d'}\bigl(R_k(t) \tilde \sigma(t_-) R_k(t)^\dagger \\
{}-\tilde \sigma(t_-)\bigr)\bigl( \rmd N_k(t) -i_k(t)\rmd t \bigr),
\end{multline}
where $\Lcal(t)$ is the following Liouville operator:
\begin{multline}\label{randomLcal}
\Lcal(t)[\tau]:=-\rmi \left[H(t),\, \tau \right] -\frac 1 2 \sum_{i=1}^d
\left\{L_i(t)^\dagger L_i(t),\tau\right\}\\ {} -\frac 1 2\sum_{k=1}^{d'} i_k(t)\left\{ R_k(t)^\dagger R_k(t),\tau \right\}\\
{}+ \sum_{i=1}^d  L_i(t) \tau L_i(t)^\dagger +\sum_{k=1}^{d'}i_k(t)R_k(t) \tau
R_k(t)^\dagger .
\end{multline}
Let us stress that this operator is random. In particular, this makes the solution
$\tilde\sigma(t)$ non-Markovian since the randomness of the operator $\Lcal(t)$ introduces a dependence on the
past. This fact will be made explicit in the concrete model developed in Sec.\ \ref{sec:model}.

Let us note that the usual master equations (without the driving noises $B$ and $N$), but with
stochastic Liouville operators, have already been considered in the literature as models of
non-Markovian evolutions. Moreover, these equations have been derived from unitary
system-environment dynamics by various techniques and approximations; see, for instance,
\cite{stochL}.

\subsection{The nonlinear SME}

Note that the probability density \eqref {newdensity} of $\Pbb^t_{\phi_0}$ with respect to $\Qbb$ can be
written as $p(t)=\Tr\{\tilde \sigma(t)\}$. Then, we normalize $\tilde\sigma(t)$ by defining the state
$\tilde\rho(t)=\tilde\sigma(t)/\Tr\{\tilde\sigma(t)\}$; when the denominator vanishes we take for
$\tilde\rho(t)$ an arbitrary state.  It is then possible to show that $\tilde\rho(t)$ satisfies the nonlinear
SME under the new probability $\Pbb^T_{\phi_0}$ \cite[Remark 3.6]{BarH95}:
\begin{multline}\label{NLSME}
\rmd \tilde \rho(t)=\Lcal(t)[\tilde \rho(t_-)]\rmd t+ \sum_{i=1}^d \bigl( L_i(t)
\tilde \rho(t_-) \\ {}+ \tilde \rho(t_-) L_i(t)^\dagger
-m_i(t)\tilde \rho(t_-)\bigr)\rmd W_i(t)\\
{}+\sum_{k=1}^{d'}\left(\frac{R_k(t) \tilde \rho(t_-) R_k(t)^\dagger
}{\Tr\{R_k(t)^\dagger R_k(t) \tilde \rho(t_-)\}}-\tilde \rho(t_-)\right)\\ {}
\times\bigl( \rmd N_k(t) -j_k(t)\rmd t \bigr).
\end{multline}
Everything can be expressed in terms of density matrices as we can write
\begin{equation*}
m_i(t)=2\RE\,\Tr\{L_i(t)\tilde\rho(t_-)\},
\end{equation*}
\[
j_k(t)=i_k(t)\Tr\{R_k(t)^\dagger R_k(t)\tilde\rho(t_-)\}.
\]

The nonlinear SME for $\tilde \rho(t)$ can also be directly obtained from
\eqref{nlSSE} by remarking that
\begin{equation}\label{tilderho}
\tilde \rho(t)=|\psi(t)\rangle \langle \psi(t)|.
\end{equation}

\subsection{The a priori states and the mean evolution}\label{sec:apriori}

The mean state, or a priori state, is defined by
\begin{equation}\label{def:eta}
\eta(t):=\Ebb_\Qbb[\tilde \sigma(t)]\equiv \Ebb_{\Pbb^T_{\phi_0}}[\tilde \rho(t)].
\end{equation}
By Eqs.\ \eqref{lSME} and \eqref{NLSME} one obtains
\begin{equation}
\dot \eta(t)=\Ebb_\Qbb\big[\Lcal(t)[\tilde \sigma(t)]\big]\equiv \Ebb_{\Pbb^T_{\phi_0}}
\big[\Lcal(t)[\tilde \rho(t)]\big].
\end{equation}
A major difference with the usual Markovian situation is that in our case this
equation is not closed. In the Markovian case one obtains
an equation of the form  $\dot \eta(t)=\Lcal(t)[\eta(t)]$, but in our situation this
is not possible, since the operator $\Lcal(t)$ is random and contributes to
the mean. Formally, a closed equation can be obtained by using projection techniques
such as the Nakajima-Zwanzig method. This construction has been derived in \cite{BDPP11}, but the
final equation is essentially not tractable.

It is then clear that the mean evolution is highly non-Markovian. It is important to
notice that our approach ensures that this evolution stays completely positive. We
then obtain a completely positive non-Markovian behavior, the memory effect being
encoded into the random Liouville operator $\Lcal(t)$. In particular when $\Lcal(t)$
is not random, we recover the usual Markovian framework.

\subsection{Measurement interpretation}

In this section, we present the essential ingredients needed in order to describe the
measurement interpretation of our theory.

\subsubsection{Observed outputs}\label{sec:output}

Let us consider $a_{\ell j}(t,s)$, $\ell=1,\ldots,m_I$, $n_{hk}(t,s)$,
$h=1,\ldots,m_J$, which are adapted and c\`{a}gl\`{a}d kernels and  $e_\ell(t)$,
$w_k(t)$ which are adapted and c\`{a}dl\`{a}g
processes. We can then define the following processes, which represent the outputs of the continuous
measurement process:
\[
I_\ell(t):= \sum_{j=1}^m\int_0^t a_{\ell j}(t,s)\,\rmd M_j(s)+ e_\ell(t), \]
\[
J_h(t):= \sum_{k=1}^{d'}\int_{(0,t]} n_{hk}(t,s)\,\rmd N_k(s)+w_k(t).
\]

The idea underlying the construction of these processes is that the instantaneous outputs are the formal derivatives
$\dot M_j(t)$ and $\dot N_k(t)$. The measuring apparatuses have a smoothing effect on the singular
instantaneous outputs and  can also provide some post-measurement processing of the outputs. These effects
are represented by the integrals with the detector response functions $ a_{\ell j}$ and $ n_{hk}$. Moreover,
it is possible that the detectors introduce some further noise, for instance of electronic origin,
and this is taken into account by the additive noises $ e_\ell$ and $ w_k$ and by the fact that response functions
can be random.

Let us consider now all the events that can be observed up to time $t$, that is, the events determined by
the outputs $I_\ell$, $J_h$ up to $t$. Let us denote by $\Gscr_t$ the collection of such events. In mathematical terms
$\Gscr_t$  is the $\sigma$-algebra generated by $I_\ell(s)$ and
$J_h(s)$, with $s\in[0,t]$, $\ell=1,\ldots,m_I$, $h=1,\ldots,m_J$. Because all the processes involved in
the definition of the outputs are $(\Fscr_t)$ adapted, we get
$\Gscr_t\subset\Fscr_t$, for all $t$. Let us stress
that in general we do not have $\Gscr_t=\Fscr_t$, because $\Gscr_t$ contains only events that can be observed by the
measuring apparatuses, while $\Fscr_t$ can contain
extra sources of noise, which can affect the system (a noisy environment for instance).

\subsubsection{Feedback} In this formalism we can describe also measurement-based feedback: parts of the outputs
are used to control some features of the dynamics or of the
measuring apparatus, say through a stimulating laser or through a local oscillator in a
homo- or heterodyne detector. When the feedback involves the output in the past, other memory
effects are introduced. A typical measurement-based feedback is represented by a Hamiltonian term functionally
dependent on some output up to the current time; while in this way it becomes a random Hamiltonian, its contribution is
perfectly compatible with the whole formalism. We shall not give examples in this paper; the theory and some
applications can be found in \cite[Sec.\ 4.4]{BarH95} and \cite{BarG11}.

\subsubsection{Instruments and a posteriori state}

A cornerstone of a consistent measurement interpretation of SME relies on the
introduction of the so-called \textit{instruments}. In order to develop this theory
we need to define the propagator $\Acal(t,s)$ of Eq.\ \eqref{lSME}, that is, the
random linear map $\tilde\sigma(s)\mapsto\tilde\sigma(t)$. An essential point is
that this application is completely positive and satisfies the composition rule
$\Acal(t,s)= \Acal(t,r) \circ\mathcal A(r,s)$, $0\leq s\leq r\leq t$.

Now for an event $A\in\Gscr_t$, we define
\begin{equation}
\Ical_t(A)[\rho]=\Ebb_\Qbb[\mathbf 1_A\mathcal{A}(t,0)[\rho]].
\end{equation}
For all $A\in \Gscr_t$, $\mathcal I_t(A)$ is a completely positive linear map called
an \textit{instrument}. In particular this gives the probability that an event
$A\in\Gscr_t$ occurs. More precisely, if $\rho_0$ represents the pre-measurement
state, the probability of $A\in\Gscr_t$ is given by
\begin{equation}
\Tr\{\mathcal I_t(A)[\rho_0]\}=\mathbb P^t_{\phi_0}[A], \qquad \Ebb_\Qbb[|\phi_0\rangle \langle \phi_0|] =\rho_0,
\end{equation}
and we recover the previous definition of the physical probability.

Then, we can define the \textit{a posteriori} state by
 \[ \rho(t):=\Ebb_{\Pbb^T_{\phi_0}}[\tilde
\rho(t)|\Gscr_t]\equiv \frac{\sigma(t)}{\Tr\{\sigma(t)\}}, \]
where
\[
\sigma(t):=\Ebb_{\Qbb}[\tilde \sigma(t)|\Gscr_t].
\]
The state $\rho(t)$ corresponds
to the update of the state of the system conditionally on the observation of the
outputs up to time $t$.

It is important to notice that in general we can not derive a closed equation for
$\rho(t)$ such as the one for $\tilde\rho(t)$. Essentially, it depends whether or not
$\Gscr_t=\Fscr_t$. In the case of $\Gscr_t \neq\Fscr_t$ the randomness of the
operators appearing in $\Lcal(t)$ will prevent the equation from being closed; again some
projection technique could be used to obtain a kind of closed equation, but it would be
intractable for practical purposes.

As a conclusion, we can see that this approach allows us to describe non-Markovian
evolutions that are generalizations of the Markovian setup.  As we shall see, the
randomness of the operators $L_i(t)$ and $R_k(t)$ will be used to describe concrete
non-Markovian effects such as colored environments and incoherent stimulating light.

The physical model is determined by the physical probability, the nonlinear SME, and
the outputs, not by the SSE, which is not unique. Two SSEs giving solutions that differ only by a
stochastic phase are physically equivalent; no physical consequence depends on a global phase in
$\phi(t)$ or $\psi(t)$ \cite[Sec.\ 2.5]{BarG09}.

\section{A model: a noisy oscillator}\label{sec:model}
Let us present now a mathematically treatable but sufficiently rich and physically
interesting model; the aim is to understand what kind of physical phenomena and
memory effects can be described by the theory we have presented. To be simple we
take a linear system, but we allow for absorption, emission, colored noises acting
on the system and on the detection apparatuses, and so on. The general scheme is the
following:
\begin{enumerate}
\item The quantum system is a single oscillator; to fix the ideas we think of a mode in an optical cavity,
but it could be an ion in a trap or some other system in the harmonic approximation.
Let $a$ and $a^\dagger$ be the usual annihilation and creation operators of quanta
in the mode; then, the free Hamiltonian of the oscillator is
\begin{equation}\label{def:H0}
H_0(t)\equiv H_0=\nu_0 a^\dagger a, \qquad \nu_0>0.
\end{equation}

\item The system emits and absorbs light; the system-electromagnetic-field interaction is treated in the usual Markov approximation.
\begin{enumerate}
\item Some emitted light reaches a photocounter: direct detection. The post-processing
of the output is taken into account by a detector response function.
\item Some light reaches a homo- or heterodyne detector. The function describing the local
oscillator can be random, a way to model imperfections. We shall show that this fact
introduces memory in the detection process, not in the mean dynamics. Moreover, we
can have also a detector response function acting as a frequency filter; see Eq.\
\eqref{Gm1}.
\item We introduce a stimulating laser; the laser-wave-oscillator interaction is treated
in the usual dipole and rotating-wave approximations. The laser wave can be random
because the laser is noisy and/or because of feedback. This introduces memory also
into the Liouvillian and in the mean dynamics, in spite of the fact that the
interaction is without memory.
\end{enumerate}
\item We introduce various kinds of colored environments. According to the choices of
the parameters these new terms can describe incoherent light, a squeezed reservoir,
a usual (or colored) thermal bath, intermediate situations, and so on.

\end{enumerate}

As in the general part, also in this model $B_1,\ldots,B_d$ are $d$ independent
standard Wiener processes under the reference probability $\Qbb$; here we shall have
$d\geq 5$. Moreover, we shall introduce $m=5$ diffusive channels and a single jump
channel, $d'=1$. Finally, we shall introduce a single diffusive output and a single
counting output; according to the notations of Sec.\ \ref{sec:output} we shall
have $m_I=m_J=1$.

\subsection{Stimulating laser and emitted light detection}

As already said we consider the oscillator-electromagnetic-field interaction in the dipole and
rotating-wave approximations. We divide the directions of the propagating light into
some ``channels''. The index 1 labels the ``side'' channels used to describe
the emitted light reaching a photo-counter (direct detection) or a
homo- or heterodyne detector. Channels 2 and 3 are the ``forward'' channels in the
direction of the stimulating laser; they describe also losses of light.

\subsubsection{Detection}

\paragraph{Direct detection.} We consider only one counter, so that we have
$d'=1$. Under the reference probability $\Qbb$, the associated counting process
$N_1(t)\equiv N(t)$ is taken to be a Poisson process of intensity $i_1(t)\equiv \lambda\geq 0$. When
$\lambda=0$ this channel is not open. The associated operator is
\begin{equation}\label{R1}
R_1(t)\equiv R= \overline{\beta }\,a, \qquad \beta\in \Cbb.
\end{equation}

With respect to the general case of Sec.\ \ref{sec:output}, let us consider only
deterministic, time invariant, real and continuous detector response functions, so
that we have
$
J_1(t)\equiv J(t)$, $n_{11}(t,s)=F_J(t-s)$, $w_1(t)=0$,
and the output current is
\begin{equation}\label{J(t)}
J(t)=\int_0^t F_J(t-r)\,\rmd N(r).
\end{equation}

\paragraph{Homodyne or heterodyne detection.}
As usual, homo- or heterodyne detection is described in the Markov approximation by a
diffusive channel driven by a Wiener process
$B_1$ (under the reference probability $\Qbb$) \cite[Sec.\ 7.2]{BarG09}. By particularizing
the quantities introduced in Sec.\ \ref{sec:SSEa}, we have $L_1^0(t)=L_1(t)$ and $ M_1(t)=B_1(t)$, which
means $f_1(t)=0$ and $b_{1i}(t)=\delta_{1i}$. Then we take
\begin{equation}\label{L1}
L_1(t)= -\rmi \overline{\alpha_1}\,\overline{h(t)}\,a, \qquad  \abs{h(t)}=1, \quad
\alpha_1\in \Cbb;
\end{equation}
$h(t)$ is the contribution of the local oscillator, which can be random. Randomness
in the local oscillator can be due to imperfections, but it could be due also to the
fact that $h(t)$ is taken dependent on some of the observed outputs at previous
times in order to describe adaptive measurements, as is done in \cite[Sec.\ 7.9.2]{WisM10}.

We consider again a deterministic, time invariant, real and continuous response
function $F_I$; in terms of the notation of Sec.\ \ref{sec:output} we take $
I_1(t)=I(t)$, $a_{1j}(t,s)=F_I(t-s)$, and $e_1(t)=0$. Then, the output current of the
homo- or heterodyne detector is
\begin{equation}\label{I(t)}
I(t)=\int_0^t F_I(t-r)\,\rmd B_1(r).
\end{equation}
We assume the response function
$F_I$ to be in $L^1(\Rbb_+)$, so that its Fourier transform exists:
\begin{equation}\label{G_I}
G_I(\mu):= \int_0^{+\infty}\rme^{\rmi \mu t}F_I(t)\,\rmd t.
\end{equation}
We shall see in Sec.\ \ref{sec:hhdet} that $\abs{G_I(\mu)}^2$ has the role of a
linear frequency filter on the output.

\paragraph{Contribution to the linear SSE.}
Summarizing, the contributions to the right hand side of the linear SSE
\eqref{lSSE1} or \eqref{lSSE2} of the two detection channels are
\begin{multline*}
\frac\lambda 2 \left( \openone - \abs{\beta}^2 a^\dagger a \right)\phi(t_-)\,\rmd t
+\left(\overline{\beta}\,a -\openone\right)\phi(t_-)\,\rmd N(t)
\\
{} -\frac{\abs{\alpha_1}^2}2 \, a^\dagger a\,\phi(t_-)\,\rmd t -\rmi
\overline{\alpha_1}\,\overline{ h(t)}\, a\phi(t_-)\,\rmd B_1(t).
\end{multline*}
The final linear SSE is given by Eq.\ \eqref{eq:lSSmodel}.

\subsubsection{The forward channels}\label{sec:fc}

Channels 2 and 3 represent the forward channel (the direction of the stimulating
laser) and the lost light; we can include in these channels other Markovian
dissipative contributions. There is no detector associated with these channels, and we
choose to put a diffusive component (the Wiener $B_2$) in channel 2, while channel 3 is
used to complete the Hamiltonian part with the contribution of the stimulating
laser. With respect to the symbols used in the linear SSE \eqref{lSSE1} and in
Sec.\ \ref{sec:SSEa} we take
\begin{equation}\label{L20}
L_2^0(t)\equiv L_2^0=-\rmi \overline{\alpha_2}\, a, \qquad L_3^0(t)\equiv L_3^0=-\rmi \alpha_2 a^\dagger,
\end{equation}
with $\alpha_2\in \Cbb$, and $\overline{f_2(t)}=f_3(t)=f(t)$, $ b_{2i}(t)=\delta_{2i}$, $b_{3i}(t)=0$, which
give $L_2(t)=L_2^0$, $L_3(t)=0$,
\begin{equation}
\rmd M_2(t)=\overline {f(t)}\rmd t + \rmd B_2(t), \qquad \rmd M_3(t)=f(t)\rmd t.
\end{equation}

\paragraph{Contribution to the linear SSE.}\label{par:SSE1}
Summarizing, the contributions to the right-hand side of the linear SSE
\eqref{lSSE2} of  channels 2 and 3 are
\[
\left( -\frac{\abs{\alpha_2}^2}2 \, a^\dagger a -\rmi H_f(t) \right)\phi(t_-)\rmd t
-\rmi\overline{\alpha_2}\, a\phi(t_-)\rmd B_2(t),
\]
where $H_f(t)$ contains the interaction between the stimulating external laser and
the oscillator:
\begin{equation}\label{Hf}
H_f(t)=\overline{\alpha_2}\, \overline{f(t)} \,a+ \alpha_2 f(t) \,a^\dagger.
\end{equation}

\paragraph{Stimulating laser.} The function $f(t)$ represents the laser wave, eventually a laser with
imperfections \cite{BarG09}. In the case of \emph{closed loop control}, the laser
wave could depend on the observed output \cite{BarG11}, but here we disregard the
possibility of feedback. Then a good model for a not perfectly coherent stimulating
laser is the phase diffusion model \cite{KM77}. Let $\nu_3>0$ be the carrier frequency of the laser light (in this case
$\Delta\nu=\nu_0-\nu_3$ is called the \emph{detuning}) and let $\varepsilon>0$ be its
bandwidth; then
 \begin{equation}\label{eq:stim.ph}
f(t)=g \exp\left\{-\rmi \nu_3 t +\rmi\sqrt{\varepsilon}\, B_{3}(t)\right\}\,,\quad
g\in\Cbb.
 \end{equation}
The quantity $g$ contains the amplitude and the initial phase of the laser; in
principle it could be a random variable, but for simplicity here we take it to be
deterministic.

To identify the bandwidth of the laser light $f$, we consider its spectrum. Since
$f$ is a complex stochastic process, its spectrum is given by the classical definition
\cite{Howard}
\begin{equation}\label{fspectrum1}
S_f(\mu):= \lim_{T\to +\infty}
\frac 1 T \,\Ebb_\Qbb \left[\abs{\int_0^T \rme^{\rmi \mu t} f(t)\,\rmd t}^2\right].
\end{equation}
By using the autocorrelation function \eqref{mom:ffbar} of the process $f$  we
easily get the Lorentzian spectrum
\begin{equation}\label{fspectrum2}
S_f(\mu)= \frac{\varepsilon
\abs{g}^2}{\left(\mu-\nu_3\right)^2+\varepsilon^2/4}.
\end{equation}

\paragraph{Homodyne detection.} In this case the local oscillator and the stimulating light
are generated by the same laser. A choice, that takes into account the differences
in the optical paths, is
\begin{equation}\label{lo_homo}
h(t)=\rme^{\rmi \theta}\,\frac{f(t-\Delta t)}{\abs{f(t- \Delta t )}}, \quad \theta\in \Rbb;
\end{equation}
we are assuming $f(t)\neq 0$. The phase $\theta$ and the time shift $ \Delta t $ depend on the physical
implementation of the homodyne apparatus and could be random, but, for simplicity,
we take both to be deterministic, $\theta \in [0,2\pi)$ and $ \Delta t \in \Rbb$.

\paragraph{Heterodyne detection.} The local oscillator $h(t)$ and the stimulating wave $f(t)$
are produced by different laser sources and the phase difference is not stable; the carrier frequencies are
generally different. In this case $h$ could depend on the output (another form of
closed loop control) or could be described by a phase diffusion model (noise in the
local oscillator). In this second case we can take
\begin{equation}\label{locosc}
h(t)=\exp\left\{\rmi \vartheta-\rmi \nu
t +\rmi\sqrt{\kappa}\, B_{4}(t)\right\},
\end{equation}
$\nu\in \Rbb$, $\kappa> 0$, $\vartheta\in \Rbb$.

\subsubsection{Summary of the contributions to the linear SME}\label{sec:SME1}
We have already explicitly given the various contributions to the SSE. To understand better the meaning of
these terms it is worthwhile to write down how they contribute to the linear SME \eqref{lSME} and to the random
Liouville operator \eqref{randomLcal}. Let us consider the free Hamiltonian of the oscillator and all the other
terms we have introduced up to now; let us set
\begin{equation}\label{gamma0}
\gamma_0 :=
\abs{\alpha_1}^2+\abs{\alpha_2}^2 +\abs{\beta}^2\lambda>0,
\end{equation}
and  $\tilde \sigma(t):=|\phi(t)\rangle \langle \phi(t)|$ as in Sec.\ \ref{sec:lSME}.
Then, we have
\begin{multline}\label{dsigma1}
\rmd \tilde \sigma(t)=\Lcal_{\mathrm{em}}(t)[\tilde \sigma(t_-)]\rmd t\\ {}+
\left(\abs{\beta}^2 a\tilde \sigma(t_-) a^\dagger -\tilde \sigma(t_-)\right)\bigl(
\rmd N(t) -\lambda\rmd t \bigr)\\
{}+\left( \rmi \alpha_1 h(t)\tilde \sigma(t_-)a^\dagger-\rmi
\overline{\alpha_1}\overline{h(t)}\,a\tilde \sigma(t_-) \right)\rmd B_1(t)
\\ {}+ \left(\rmi \alpha_2 \tilde
\sigma(t_-)a^\dagger-\rmi \overline{\alpha_2}\,a\tilde \sigma(t_-)\right)\rmd
B_2(t)+\cdots
\end{multline}
(the ellipsis stands for further contributions that we shall introduce in Sec.\ \ref{sec:col}),
\begin{multline}\label{Lem}
\Lcal_{\mathrm{em}}(t)[\tau]=-\rmi [H_0+H_f(t),\, \tau ]
\\ {}+\gamma_0 a \tau a^\dagger-\frac{\gamma_0} 2\left\{a^\dagger a , \tau\right\}.
\end{multline}

From these equations it is apparent that the electromagnetic interaction has been
treated in the usual Markov approximation; we see also that the parameter $\gamma_0$
is the mode width. The only possible sources of memory are $f$ (the stimulating
laser light) and $h$ (the local oscillator). So, up to now the memory is due only to
the imperfections inducing randomness in the lasers involved. Remember that we have
not included a conceptually very important source of memory, the possibility of
feedback.

\subsection{A colored environment}\label{sec:col}

Our aim here is to introduce some sources of colored noise; they could describe
physically different scenarios, which we shall discuss at the end of the section.

Let us introduce a complex Gaussian process $Y$ given by
\begin{equation}\label{Y(t)}
Y(t):=\sum_{j=5}^d \left(b_j B_j(t)+\int_0^tX_j(s)\,\rmd s \right),
\end{equation}
\begin{equation}\label{Xj(t)}
X_j(s)=\int_0^s   c_j(s-u)\,\rmd B_j(u).
\end{equation}
Here $b_j\in\Cbb$ and we set
\begin{equation}\label{qk}
q:=\sum_{j=5}^d {b_j}^2, \qquad k:=\sum_{j=5}^d
\abs{b_j}^2;
\end{equation}
moreover, we assume the complex functions $c_j$ to be integrable, i.e.
\begin{equation}\label{assumption1}
\int_0^{+\infty}\abs{c_j(t)}\rmd t<+\infty.
\end{equation}

Now, we add two more diffusive channels; with the notations of Sec.\ \ref{sec:SSE}, we
take $m=5$ and
\begin{equation}
L_4^0(t)\equiv L_4^0= -\rmi a, \qquad L_5^0(t)\equiv L_5^0= -\rmi a^\dagger,
\end{equation}
\begin{equation}
M_4(t)=\overline{Y(t)}, \qquad M_5(t)=Y(t),
\end{equation}
that means
$
\overline{f_4(t)}=f_5(t)=\sum_{j=5}^d X_j(t)$,
\[
\overline{b_{4i}(t)}= b_{5i}(t) =\begin{cases}0, & i\leq 4, \\ b_i, & i\geq
5\end{cases}.
\]
This gives $L_4(t)=0$ and, for $i\geq 5$,
\begin{equation}\label{L>4}
L_i(t)\equiv L_i= \overline{b_i}\, L_4^0+ b_i L_5^0=-\rmi \left(\overline{b_i}\,a+b_i a^\dagger\right).
\end{equation}

The contribution of these new terms to the linear SSE \eqref{lSSE1} turns out to be
\begin{equation}\label{contrSSE2}
-\frac {D}
2\,\phi(t_-)\rmd t-\rmi \left( a^\dagger\rmd Y(t)+a \rmd \overline{Y(t)}\right)
\phi(t_-),
\end{equation}
where
\begin{equation}\label{operatorD}
D:=2ka^\dagger a +k+\overline{q}\,a^2
+ q\,{a^\dagger}^2.
\end{equation}

\subsubsection{The spectrum of the Gaussian noise}

The dynamics of our system involves the differential of the process $Y$ or, in other
terms, its generalized derivative $\dot Y(t)$. Like the spectrum of $f$ \eqref
{fspectrum1}, the spectrum of this classical complex process is defined by
\begin{equation}\label{Yspectrum1}
S_Y(\mu):= \lim_{T\to +\infty}
\frac 1 T \,\Ebb_\Qbb \left[\abs{\int_0^T \rme^{\rmi \mu t} \rmd Y(t)}^2\right],
\end{equation}
when the limit exists.

By construction, $Y$ is a Gaussian process with zero mean; its second moments,
needed in \eqref{Yspectrum1}, can be easily computed by using the properties of the
stochastic integrals (the It\^o isometry).

Let us introduce the Laplace transform of $c_j$
\begin{equation}\label{LapTc}
s_j(z):=\int_0^{+\infty}\rme^{-z t} c_j(t)\rmd t, \qquad \RE z\geq 0,
\end{equation}
which exists owing to the integrability condition \eqref{assumption1}.

The spectrum \eqref{Yspectrum1} is computed in Appendix \ref{A4} and it is given by
\begin{equation}\label{Yspectrum2}
S_Y(\mu)=
\sum_{j=5}^d S_{Y_j}(\mu), \quad S_{Y_j}(\mu)=
\abs{b_j+s_j(-\rmi\mu)}^2.
\end{equation}
Note that the spectrum of $Y(t)$ is the sum of the spectra of the components $Y_j(t)$ without any
interference among them. Each spectral component contains a white-noise contribution ($b_j$) and a regular
one ($s_j$), which interfere (they sum up inside the square modulus).
Moreover, let us stress that by this construction it is possible to insert Gaussian
noises with given spectra, not only in this model, but even in the general theory.

\subsubsection{Contribution to the linear SME}
As was done for the electromagnetic contributions in Eq.\ \eqref{dsigma1}, it is useful
to identify the contributions to the linear SME due to the new noises:
\begin{multline}\label{dsigma2}
\rmd \tilde \sigma(t)=\cdots+ \sum_{j=5}^d\Bigl(\Lcal_{j}(t)[\tilde \sigma(t_-)]\rmd t\\
{}-\rmi \left[C_j,\,\tilde \sigma(t_-)\right]\rmd B_j(t)\Bigr)
\end{multline}
(the ellipsis stands for the contributions already introduced), where
\begin{subequations}\label{LG}
\begin{equation}
\Lcal_{j}(t)[\tau]:=-\rmi\left[H_j(t) ,\, \tau \right]
-\frac 1 2  \left[C_j , \left[C_j,\, \tau \right]\right],
\end{equation}
\begin{equation}\label{Hj}
C_j:=b_j a^\dagger +\overline{b_j}\, a ,\qquad H_j(t):=X_j(t) a^\dagger+\overline{X_j(t)}\,a.
\end{equation}
\end{subequations}

Let us note that the contributions of the classical processes $X_j$ to the dynamics \eqref{dsigma2}
are very reminiscent of the contribution of classical processes in the ``adjoint equation'' in
\cite[Sec.\ 3.5]{GarZ04}. This shows that these contributions, or at least some of them, can come
from the interaction of the system with a quantum reservoir and could be derived by using the
techniques of the quantum Langevin equation and the adjoint equation \cite[Secs.\ 3.1 and 3.5]{GarZ04}.

Now we can identify the physical meaning of various possible  contributions.

\paragraph{Incoherent light.}\label{par:il}
Consider the index $j=5$ and assume $b_5=0$. Then, $C_5=0$ and this term contributes
only with a regular random Hamiltonian term $H_5(t)$. Its structure is very similar
to that of $H_f(t)$, but the two random processes involved are qualitatively different. The
process $f$ is the exponential of a Gaussian process and represents a
quasi-monochromatic wave (laser light). The process $X_5$ is Gaussian and could
represent, for instance, incoherent light with an arbitrary spectrum $\abs{s_5(-\rmi
\mu)}^2$, for instance, thermal light with a black-body spectrum \cite[Eqs.\ (1.52) and (7.148)]{Carm02}
\[
\abs{s_5(-\rmi\mu)}^2= \frac{6\hbar^2\mu \rme^{-\hbar \mu/k_B T}}{\pi^2 k_B^{\, 2} T^2
\left(1-\rme^{-\hbar \mu/k_B T}\right)}, \qquad \mu>0.
\]

Another possible choice for incoherent light is an Ornstein-Uhlenbeck process, that
means taking
\[
c_5(t)=1_{(0,+\infty)}(t) g_5 \rme^{-\overline{\varkappa_5}\, t},
\qquad \varkappa_5=\frac{\gamma_5}2-\rmi \nu_5,
\]
with $g_5\in \Cbb$, $\nu_5\in \Rbb$, $\gamma_5>0$. In this case, from
\eqref{LapTc} we get
\[
s_5(z)=\frac{g_5}{z+\overline{\varkappa_5}},
\]
and the contribution to the spectrum \eqref{Yspectrum2} is the Lorentzian term
\begin{equation}\label{s5}
\abs{s_5(-\rmi\mu)}^2=\frac{\abs{g_5}^2}{\left(\mu-\nu_5\right)^2+{\gamma_5}^2/4}.
\end{equation}
As already seen, also the phase diffusion model \eqref{eq:stim.ph} of the laser  gives a Lorentzian spectrum
\eqref{fspectrum2}, but, in spite of this, the two cases are completely different. The wave
\eqref{eq:stim.ph} is quasi-coherent, while the Ornstein-Uhlenbeck process represents a Gaussian incoherent wave.

\paragraph{Squeezed reservoir.} \label{par:sqR}
Consider now the indices $j=6,7$ and assume
$c_{6}= c_{7}=0$; then, we get $X_6=X_7=0$ and the contributions of these terms are Markovian. Indeed, by defining
\begin{equation}\label{nm}
n:=\frac{\abs{b_{6}}^2 +\abs{b_{7}}^2}{\gamma_0}\,, \qquad m:=
-\frac{{b_{6}}^2 +{b_{7}}^2}{\gamma_0}\,,
\end{equation}
\[
\rmd C(t):= \frac{b_6\, \rmd B_6(t)+b_7\,\rmd B_7(t)}{\sqrt{\abs{b_{6}}^2 +\abs{b_{7}}^2}}\,,
\]
we get
\begin{multline*}
\sum_{j=6}^7\Lcal_{j}(t)[\tau]=\gamma_0 n \biggl(a\tau a^\dagger +a^\dagger\tau a
-\frac 1 2 \bigl\{a^\dagger a\\ {}+a a^\dagger,\,
\tau\bigr\} \biggr)- \gamma_0  m\left(a^\dagger\tau a^\dagger -\frac 1 2 \left\{a^\dagger
a^\dagger, \tau\right\}\right)\\ {}-\gamma_0 \,\overline{m} \left(a\tau a -\frac 1 2 \left\{a a,
\tau\right\}\right),
\end{multline*}
\begin{multline*}
-\rmi \sum_{j=6}^7 [C_j,\tau]\rmd B_j(t)\\ {}=\rmi \sqrt{\gamma_0 n} \Bigl( [\tau,a]\rmd \overline{C(t)}
-[a^\dagger,\tau]\rmd C(t)\Bigr).
\end{multline*}

By combining these contributions with the dissipative term in $
\Lcal_{\mathrm{em}}(t)$ \eqref{Lem} we get the typical dissipative effect of a squeezed reservoir
\cite[Eq.\ (10.2.42)]{GarZ04}. So we can interpret $n$ as the effective photon number and $m$ as
the squeezing parameter of the reservoir.
When $m=0$, i.e. $b_{7}=\pm \rmi b_{6}$, the previous dissipative terms reduce to the contribution of a
\emph{thermal bath} with $\gamma_0n=2\abs{b_{6}}^2$.

\paragraph{The generic case.} The spectrum \eqref{Yspectrum2} shows that the generic case is in between the two
cases discussed above: a Markovian dissipative contribution and a random Hamiltonian contribution with interference
between them.

\subsection{The full model}
Putting together all the contributions we have introduced in Eqs.\ \eqref{def:H0}, \eqref{contrSSE2} and
in Sec.\ \ref{par:SSE1}, we get that the full linear SSE can be written as
\begin{multline}\label{eq:lSSmodel}
\rmd \phi(t)=\left(-\overline{\varkappa_0}  a^\dagger a +\frac {\lambda -D} 2\right)
\phi(t_-)\rmd t \\
{}-\rmi a^\dagger\phi(t_-)\rmd Y_1(t)-\rmi a\phi(t_-)\rmd Y_2(t)
\\ {}+ \bigl(\overline{\beta}\,
a-\openone\bigr)\phi(t_-)\,\rmd N(t),
\end{multline}
where $\lambda$ is the intensity of the Poisson process $N$,
\begin{subequations}
\begin{equation}
\rmd Y_1(t)=\alpha_2 f(t)\rmd t +\rmd Y(t),
\end{equation}
\begin{multline}
\rmd Y_2(t)=\overline{\alpha_1} \, \overline{h(t)}\,\rmd B_1(t)
+\overline{\alpha_2} \, \overline{f(t)}\,\rmd t \\ {}+\overline{\alpha_2} \,\rmd
B_2(t) +\rmd \overline{Y(t)},
\end{multline}
\end{subequations}
\begin{equation} \label{kappa0}
\varkappa_0=-\rmi \nu_0+\frac {\gamma_0 } 2, \qquad \abs{h(t)}=1;
\end{equation}
$\gamma_0$,  $Y$, and $D$,
are given by Eqs.\ \eqref{gamma0}, \eqref{Y(t)}, and \eqref{operatorD}.

\subsubsection{The Liouville operator and the linear SME}\label{sec:Liouv_model}

From Eqs.\ \eqref{Lem} and \eqref{LG} we get the full
Liouville operator
\begin{multline}
\Lcal(t)[\tau]=-\rmi [H(t),\, \tau ]+\gamma_0 \left( a\tau a^\dagger - \frac 1 2 \left\{ a^\dagger
a ,\, \tau \right\}\right)
\\
{}-\frac 1 2  \sum_{j=5}^d\left[ b_j a^\dagger +\overline{b_j}\, a , \left[b_j a^\dagger
+\overline{b_j}\, a ,\, \tau \right]\right],
\end{multline}
\begin{equation*}
H(t)=H_0+\overline{\mathtt{f}(t)}\, a
+\mathtt{f}(t) a^\dagger,
\end{equation*}
\begin{equation*}
\mathtt{f}(t)=\alpha_2 f(t)+\sum_{j=5}^d X_j(t).
\end{equation*}
Note that the Liouville operator is random only because of the presence of the process $\mathtt{f}(t)$, but
this is enough to preclude having a closed master equation for the \emph{a priori} states, see Sec.\
\ref{sec:apriori}. Let us stress again the different physical roles of the wave $f$ and the Gaussian processes
$X_j$, as discussed in Sec.\ \ref{par:il}.

By putting together Eqs.\  \eqref{dsigma1} and \eqref{dsigma2} we get the full linear SME
\begin{multline}
\rmd \tilde \sigma(t)=\Lcal(t)[\tilde \sigma(t_-)]\rmd t-\rmi \sum_{i=5}^d \left[
\overline{b_i}\, a + b_i a^\dagger,\, \tilde \sigma(t)\right]\rmd B_i(t)
\\ {}-\rmi
\left(\overline{\alpha_1}\, \overline{h(t)}\, a \tilde \sigma(t) - \alpha_1 h(t)
\tilde \sigma(t)a^\dagger\right) \rmd B_1(t)
\\ {}-\rmi
\left(\overline{\alpha_2}\,  a \tilde \sigma(t) - \alpha_2 \tilde
\sigma(t)a^\dagger\right) \rmd B_2(t)
\\
{}+\left(\abs{\beta}^2a \tilde \sigma(t_-) a^\dagger -\tilde
\sigma(t_-)\right)\bigl( \rmd N(t) -\lambda\rmd t \bigr).
\end{multline}
In the measuring process there is one more source of memory coming in through the
randomness in the local oscillator $h(t)$.

\subsubsection{The solution of the linear SSE}\label{sec:solution}
The simplifying mathematical feature of the model we have constructed is that the SSE leaves invariant the
coherent states. Such states are defined by
$
a e(\xi)=\xi e(\xi)$, $ \norm{ e(\xi)}=1$. We assume the initial condition to be the coherent vector
\[
\phi_0=
e\big(\xi_0\big), \qquad
\xi_0\in \Cbb.
\]

To find the solution of the SSE \eqref{eq:lSSmodel}
we make the ansatz
\begin{equation}\label{solution}
\phi(t)=V(t)\rme^{\frac 1 2 \,Z(t)} e\big(\xi(t)\big).
\end{equation}
As $\norm{\phi(t)}^2\equiv \abs{V(t)}^2 \rme^{\RE Z(t)}$ must be a martingale, we ask
\begin{itemize}
\item $V(t)$ to contain only the contributions from the jumps of $N(t)$ and to be such that $\abs{V(t)}^2$
is a martingale,
\item $Z(t)$ to contain only the contributions from the diffusive processes and to be such that
$\rme^{\RE Z(t)}$ is a martingale,
\item $\xi(t)$ to be a generic stochastic process whose differential contains all the possible terms.
\end{itemize}
Then, we identify $\rmd \langle e(\epsilon)|\phi(t)\rangle$ with $ \langle e(\epsilon)|\rmd\phi(t)\rangle$,
where now $\rmd\phi(t)$ is taken from Eq.\ \eqref{eq:lSSmodel}. By equating the coefficients of the monomials
with homogeneous powers of $\epsilon$ we get stochastic equations for the unknown terms, which can be solved.
The final result is
\begin{equation}\label{xi1}
\xi(t) =\rme^{-\overline{\varkappa_0} t }\xi_0-\rmi U_f(t)-\rmi U_Y(t),
\end{equation}
\begin{equation}\label{Uf}
U_f(t):=\alpha_2\int_0^t
\rme^{-\overline{\varkappa_0} (t-r) } f(r)\, \rmd r,
\end{equation}
\begin{equation}\label{UY}
U_Y(t):=\sum_{j=5}^d\int_0^t g_j(t-r)\rmd B_j(r),
\end{equation}
\begin{equation}\label{xi2}
g_j(t):=\rme^{-\overline{\varkappa_0}\, t} b_j+ \int_0^{t} \rme^{-\overline{\varkappa_0}
\left(t-u\right)}c_j(u)\,\rmd u ,
\end{equation}
\begin{multline}\label{Z(t)}
Z(t)=
-2\rmi \int_0^t\xi(s)\left(\overline{\alpha_1}\, \overline{h(s)}\, \rmd B_1(s)+
\overline{\alpha_2}\, \rmd B_2(s)\right)
\\
{}- 2\rmi\RE \int_0^t \overline{\xi(s)}\,\rmd Y_1(s)+\int_0^t\left(\overline{\alpha_1}^2 \overline{h(s)}^2+
\overline{\alpha_2}^2\right)\xi(s)^2\rmd s
\\ {}-\left(\abs{\alpha_1}^2 + \abs{\alpha_2}^2\right)\int_0^t
\abs{\xi(s)}^2\,\rmd s,
\end{multline}
\begin{multline}\label{V(t)}
V(t)=\exp\left\{\frac \lambda 2\int_0^t
\left(1-\abs{\beta}^2\abs{\xi(s)}^2\right)\rmd s\right\}
\\ {}\times\prod_{r\in(0,t]}\left[\overline{\beta}\xi(r)\right]^{\Delta N(r)}.
\end{multline}
Let us stress that, almost surely, the product in \eqref{V(t)} contains a finite number of factors different
from 1 and that the expression \eqref{V(t)}  is nothing but the solution of the linear SDE
\begin{multline*}
\rmd V(t)=V(t)\biggl\{\frac \lambda 2 \left(1-\abs{\beta}^2
\abs{\xi(t)}^2\right)\rmd t \\ {}+ \left(\overline{\beta} \, \xi(t)- 1\right)\rmd
N(t)\biggr\}
\end{multline*}
with initial condition $V(0)=1$. Independently of the methods used to find the solution, by using stochastic
calculus, one can check that the expression for $\phi(t)$ defined by \eqref{solution}-\eqref{V(t)} indeed
solves the SSE \eqref{eq:lSSmodel}.

Note that the processes $\xi(t)$ and $Z(t)$ are continuous in time and that the contribution of the jumps is concentrated in $V(t)$.

\subsubsection{The physical probability}
Let us recall that the new probability is $\Pbb^T_{\phi_0}(\rmd \omega)=p(t,\omega)\Qbb(\rmd \omega) $
\eqref{newprob}. Having the explicit form of the solution of the linear SSE, we can compute the probability
density $p(t)$ \eqref{newdensity} and the normalized vector \eqref{def:psi}, that solves the nonlinear
SSE \eqref{nlSSE}. From Eqs.\ \eqref{newdensity}--\eqref{jk} we get
\begin{equation}
p(t)=\norm{\phi(t)}^2= \abs{V(t)}^2 \rme^{\RE Z(t)},
\end{equation}
\begin{equation}\label{psixi}
\psi(t)=\exp\left\{\rmi \arg V(t)+\frac \rmi 2 \,\IM Z(t)\right\}e\big(\xi(t)\big),
\end{equation}
\[
\abs{V(t)}^2=\exp\biggl\{\int_0^t\left(\lambda -
j(s)\right)\rmd s\biggr\} \prod_{r\in (0,
t]}\left[\frac{j(r)}\lambda\right]^{\Delta N(r)},
\]
\begin{equation*}
\RE Z(t)=\sum_{i=1}^2\int_0^t\left( m_i(s) \, \rmd B_i(s)- \frac 1 2 \, m_i(s)^2\rmd
s\right),
\end{equation*}
\begin{subequations}\label{jmm}
\begin{gather}\label{m1}
m_1(t)= 2\IM\left( \overline{\alpha_1}\, \overline{h(t)}\, \xi(t)\right),
\\ \label{m2j}
m_2(t)= 2\IM\left(\overline{\alpha_2}\, \xi(t)\right),
\qquad
j(t)=\lambda\abs{\beta}^2 \abs{\xi(t)}^2.
\end{gather}
\end{subequations}
Moreover, in Sec.\ \ref{sec:fc} we find $L_3=0$ and in Sec.\ \ref{sec:col} $L_4(t)=0$ and $L_i$, $i\geq 5$,
anti-selfadjoint \eqref{L>4}. By the expression \eqref{mi} for $m_i$, this gives $m_i(t)=0$ for $i=3,\ldots,d$.

\paragraph{Girsanov transformation.}\label{par:Gt} As discussed in Sec.\ \ref{par:Gir},
under the physical probability $\Pbb^T_{\phi_0}$ the process $N(t)$ is a counting process of intensity
$j(t)$ and $(W_1,W_2,B_3,\ldots,B_d)$ is a standard Wiener process and, in particular, its components are
independent; the first two components are defined by Eq.\ \eqref{newW}.

\paragraph{Assumption on $f$ and $h$.}\label{par:fh} We assume the phase diffusion model for the
stimulating laser without feedback, so that the laser wave $f(t)$ is given by Eq.\ \eqref{eq:stim.ph}
and depends only on $B_3$. Similarly we assume the local oscillator $h(t)$ to be of the form \eqref{lo_homo}
or \eqref{locosc}, without feedback.

\paragraph{Stochastic independence.}\label{par:distrib}
The Girsanov transformation of Sec.\ \ref{par:Gt}, the assumptions of Sec. \ref{par:fh}, and the results
expressed by Eqs.\ \eqref{xi1}-\eqref{xi2} and \eqref{jmm} have important consequences. First of all,
under the physical probability, $(W_1,W_2)$ is independent of $(B_3,\ldots,B_d)$ and $\xi_0$,
which is $\Fscr_0$-measurable, and, so, $(W_1,W_2)$ is independent of $(\xi,h,f,m_1,m_2,j)$.
Then, because $(B_3,\ldots,B_d)$ is a Wiener process below the reference probability and the physical probability,
the law of $(f,h,U_f,U_Y,m_1,m_2,j)$ is the same below $\Qbb$ and $\Pbb^T_{\phi_0}$. Let us stress that
these properties are specific to the linear model constructed in this section, not to the general theory of
Secs.\ \ref{sec:SSE} and \ref{sec:sme}.

\paragraph{The a priori state.} From Eqs.\ \eqref{tilderho}, \eqref{def:eta}, and \eqref{psixi} and the properties discussed in Sec.\ \ref{par:distrib}, we get that the \emph{a priori} states are given by
\begin{equation*}
\eta(t)= \Ebb_{\Pbb^T_{\phi_0}}\left[|\psi(t)\rangle\langle \psi(t)|\right]=\Ebb_{\Qbb}\left[|e\big(\xi(t)\big)\rangle \langle e\big(\xi(t)\big)|\right].
\end{equation*}
Note that $\eta(t)$ is a classical mixture of coherent states. Indeed, to keep the example simple, we introduced only interactions leaving invariant such a class of states.

\paragraph{The initial condition.} In the next two sections we shall study the detection outputs under
the physical probability $\Pbb^T_{\phi_0}$. As we shall be interested only in the long-time behavior, no result will depend on the initial condition and
from now on we take
\begin{equation}
\xi_0=0.
\end{equation}

\subsection{Direct detection}\label{sec:Ddetection}

The output current of direct detection is $J(t)$, given by Eq.\ \eqref{J(t)}, where $N$ is a counting
process of stochastic intensity $j(t)$ \eqref{m2j}:
\begin{equation}
j(t)=\lambda \abs{\beta}^2 \norm{a \psi(t)}^2=\lambda \abs{\beta}^2 \abs{\xi(t)}^2.
\end{equation}

Let $(\Dscr_t)$ be the natural filtration of the processes $f(\cdot)$ and $Y(\cdot)$
and set $\Dscr:= \bigvee_{t\geq 0}\Dscr_t$. The stochastic intensity $j(t)$ of the counting
process $N$ is due only to the presence of $f$ and $Y$ in $\xi$. Therefore, conditionally on
$\Dscr$, $N$ is a time-inhomogeneous Poisson process. But, the law of the process $j$ is the same under
the physical probability and under the reference probability; this gives the result that the
characteristic functional of $N$ is
\begin{multline}\label{PhiN}
\Phi^N_T[k]:=\Ebb_{\Pbb^T_{\phi_0}}\left[\exp\left\{ \rmi \int_0^Tk(t)\rmd N(t)\right\}\right]
\\ {}=\Ebb_{\Qbb}\left[\exp\left\{ \int_0^T\left(\rme^{\rmi k(t)}-1\right)j(t)\rmd
t\right\}\right].
\end{multline}
Similarly, one can identify all the exclusive probability densities. By writing
\begin{multline*}
\Phi^N_T[k]=P_T(0)+ \sum_{n=1}^\infty \int_0^T\rmd t_n\int_0^{t_n}\rmd t_{n-1} \\ \cdots \int_0^{t_2}\rmd t_1\,
\exp\left\{\rmi \sum_{i=1}^n k(t_i)\right\}p_T(t_n,\ldots, t_1),
\end{multline*}
we have that the probability of no counts up to time $T$ is
\[
P_T(0)=\Ebb_\Qbb\left[\rme^{-\int_0^T j(t)\, \rmd t}\right]
\]
and the probability density of a count around time $t_1$, \ldots, a count around time $t_n$ and no other count
in between is
\[
p_T(t_n,\ldots, t_1)=\Ebb_\Qbb\left[j(t_n)\cdots j(t_1)\,\rme^{-\int_0^T j(t)\, \rmd t}\right].
\]

From Eq.\ \eqref{PhiN} one obtains also the expressions for all the multi-time correlation functions
of the processes $N$ and $J$. In particular we get
\begin{equation}\label{meanN}
\Ebb_{\Pbb^T_{\phi_0}}\left[\int_0^Tk(t)\rmd N(t)\right]
=\int_0^T\rmd t\, k(t)\Ebb_{\Qbb}\left[j(t)\right],
\end{equation}
\begin{multline}\label{2momN}
\Ebb_{\Pbb^T_{\phi_0}}\left[\int_0^Tk_1(t)\rmd N(t)\int_0^Tk_2(s)\rmd N(s)\right]
\\ {}=\int_0^T\rmd t\, k_1(t)k_2(t)\Ebb_{\Qbb}\left[j(t)\right] \\ {}+ \int_0^T\rmd
t\int_0^T\rmd s\, k_1(t)k_2(s)\Ebb_{\Qbb}\left[j(t)j(s)\right].
\end{multline}

\subsubsection{The mean counting intensity}\label{sec:Mint}

By taking the expressions \eqref{J(t)} and \eqref{m2j} for the output current $J$ and for the stochastic
intensity $j$, we get
\[
\Ebb_{\Pbb^t_{\phi_0}}\left[J(t)\right]=\lambda\abs{\beta}^2\int_0^t\rmd r\, F_J(t-r)
\Ebb_\Qbb\left[\abs{\xi(r)}^2\right].
\]
As we show in Appendix \ref{mci}, the mean of $\abs{\xi(t)}^2$ has a limit $\Lambda$ for large times and,
by taking the response function $F_J$ such that $\int_0^{+\infty}F_J(t)\,\rmd t=1$, we get
\begin{equation}\label{Dintensity}
\lim_{t\to +\infty}\Ebb_{\Pbb^t_{\phi_0}}\left[J(t)\right]=\lim_{t\to +\infty}\Ebb_\Qbb[j(t)]
=\lambda\abs{\beta}^2\Lambda,
\end{equation}
where
\begin{equation}\label{Lambda}
\Lambda=\Lambda_f+
\sum_{j=5}^d \Lambda_j\,,
\end{equation}
\begin{equation}\label{Lambdaf}
\Lambda_f=\frac{\abs{\alpha_2}^2\abs{g}^2 \left(\gamma_0+\varepsilon
\right)}{\gamma_0\left( \frac{\left(\gamma_0
+\varepsilon\right)^2}4+\left(\Delta\nu\right)^2\right)}\,,
\end{equation}
\begin{equation}\label{Lambdaj}
\Lambda_j=\frac
{1}{2\pi} \int_{-\infty}^{+\infty}  \frac
{\abs{b_j+s_j(-\rmi x)}^2}{\abs{\varkappa_0+\rmi x}^2}\,\rmd x ;
\end{equation}
$s_j$ is the Laplace transform \eqref{LapTc}.

\paragraph{Incoherent light.} In the case of incoherent light with the Lorentzian spectrum \eqref{s5},
described in Sec.\ \ref{par:il},
the contribution to $\Lambda$ is
\begin{equation}\label{Lambda5}
\Lambda_5=
\frac{\abs{g_5}^2\left(\gamma_0+\gamma_5\right)}{\gamma_0\gamma_5\abs{\varkappa_0+\overline{\varkappa_5}}^2}
;
\end{equation}
note that it has same form as the contribution of $f$.

\paragraph{Squeezed reservoir.} In the case of the Markovian squeezed reservoir of Sec.\ \ref{par:sqR}
we get $s_6=s_7=0$ and
\begin{equation}\label{Lambda67}
\Lambda_6+\Lambda_7=n,
\end{equation}
where the parameter $n$ is defined in Eq.\ \eqref{nm}. Note that the squeezing parameter $m$ does not
contribute to $\Lambda$.

\subsubsection{The Mandel $Q$-parameter}
To compute the full statistics of the counts, say all the exclusive probability densities, is not easy.
A simple, significant parameter related to a counting statistics is the Mandel $Q$-parameter defined,
for $t_0\geq 0$ and $t>0$, by
\begin{equation}\label{Qt0t}
Q_{t_0}(t):=\frac{\Var_{\Pbb^{t+t_0}_{\phi_0}}[N(t+t_0)-N(t_0)]}{
\Ebb_{\Pbb^{t+t_0}_{\phi_0}}[N(t+t_0)-N(t_0)]}-1.
\end{equation}
Because for a Poisson process this parameter is zero, in quantum optics it is usual to say that in the case of
a positive $Q$ parameter one has super-Poissonian light and sub-Poissonian light in the other case.
Sub-Poissonian light is considered an indication of non-classical effects.
By Eqs.\ \eqref{meanN} and \eqref{2momN} we get
\begin{equation}\label{Qt0t2}
Q_{t_0}(t)=\frac{\Var_{\Qbb}\left[\int_{t_0}^{t+t_0}j(s) \rmd s
\right]}{\Ebb_{\Qbb}\left[\int_{t_0}^{t+t_0}j(s) \rmd s \right]}\geq 0.
\end{equation}
The emitted light is always super-Poissonian; indeed, it is well established that
non-classical light can not be obtained in a linear system.

For large times we can define
\begin{equation}\label{Qt}
Q(t):=\lim_{t_0\to + \infty}Q_{t_0}(t).
\end{equation}
This parameter is studied in Appendix \ref{app:Qpar}; its general expression is given by Eq.\ \eqref{Qtgen}.
We have that $Q(t)$ is non decreasing, $Q(0)=0$, and the limit $\lim_{t\to + \infty}Q(t)$ exists.
The expression \eqref{Qtgen} for $Q(t)$ shows that the contributions of $f$ and $Y$ interfere and the same
holds for the various components of the process $Y$: the situation for the Mandel $Q$-parameter is much more
complex than for the mean intensity $\Lambda$, to which each channel contributes independently from the others,
as seen in Sec. \ref{sec:Mint}.

\subsection{Heterodyne and homodyne detection}\label{sec:hhdet}

In terms of the new Wiener process, the hetero- or homodyne current $I(t)$ \eqref{I(t)} can be written as
\begin{subequations}\label{I+m1}
\begin{equation}\label{hhI(t)}
I(t)=\int_0^t F_I(t-r) [ m_1(r) \rmd r + \rmd W_1(r)],
\end{equation}
\begin{multline}
m_1(t)=2\RE \left(-\rmi \overline{\alpha_1}\,\overline{h(t)} \langle \psi(t)|a \psi(t) \rangle \right)\\ {} =2\IM \left( \overline{\alpha_1}\,\overline{h(t)}  \xi(t) \right).
\end{multline}
\end{subequations}
As discussed in Sec.\ \ref{par:distrib}, in our linear model
the processes $W_1$ and $m_1$ turn out to be stochastically independent under the physical probability, so that,
modulo the detector response function $F_I$, we can interpret the output \eqref{hhI(t)} as signal, $m_1$,
plus independent white noise, $\dot W_1$.

By using Eqs.\ \eqref{xi1}-\eqref{xi2} and \eqref{I+m1} and the fact that $U_Y$ has zero mean and is independent
of $h$, we get the expression for the mean output current
\begin{multline}\label{moc}
\Ebb_{\Pbb^t_{\phi_0}}[I(t)]=-2\RE\, \overline{\alpha_1} \, \alpha_2 \int_0^t \rmd s\int_0^s \rmd
r \,F_I(t-s)\\ {}\times\rme^{-\overline{\varkappa_0} (s-r)}\, \Ebb_\Qbb [
\overline{h(s)}\, f(r)].
\end{multline}

Let us study now the spectrum of the hetero- or homodyne output, which can be obtained experimentally by a spectrum
analyzer. The current $I(t)$ \eqref{hhI(t)} is a classical stochastic process and again its spectrum is given by
the classical definition \cite{Howard}
\begin{equation}\label{def:sp}
S_I(\mu)=\lim_{T\to +\infty} \frac 1 T\, \Ebb_{\Pbb_{\phi_0}^T}\left[\abs{\int_0^T
\rme^{\rmi \mu t}I(t)\rmd t}^2\right].
\end{equation}
In Appendix \ref{app:sp} we show that we have
\begin{equation}\label{Gm1}
S_I(\mu)=\abs{G_I(\mu)}^2 \left[1+S_m(\mu)\right],
\end{equation}
where $G_I(\mu)$ is the Fourier transform \eqref{G_I} of the detector response function $F_I(t)$, $1$ is the
constant contribution of the white noise $\dot W_1$ and
\begin{equation}\label{Gm2}
S_m(\mu)=\lim_{T\to +\infty}\frac
{1} {T}\,\Ebb_\Qbb\left[\abs{ \int_0^Tm_1(t)\rme^{\rmi\mu t}\rmd
t}^2\right].
\end{equation}

From the expression \eqref{Gm1} we see that the response function $F_I$ acts as a frequency filter
$\abs{G_I(\mu)}^2$; for instance, we can take any band-pass filter. Moreover, we have always
$S_I(\mu)\geq \abs{G_I(\mu)}^2$. This is due to the absence of correlations between $W_1$ and $m_1$ and
it is interpreted as absence of squeezing in the quadratures of the emitted light, in agreement with the
fact that linear systems cannot generate non-classical light. In contrast, a nonlinear system, such as
a two-level atom, can generate squeezed fluorescent light and this can be treated by the formalism of SSE
and continuous measurements \cite{BarG09,BarG11}.

As discussed in Appendix \ref{app:sp}, the spectrum $S_m(\mu)$ can be computed, and we obtain
\begin{equation}\label{Gm3}
S_m(\mu)=S_{11}(\mu)+ S_{12}(\mu)+ S_{2}(\nu_4+\mu)+S_{2}(\nu_4-\mu),
\end{equation}
where the first two contributions are given by Eqs.\ \eqref{S11} and \eqref{S12} and
\begin{equation}\label{eq:S2}
S_{2}(\mu)=
\int_{-\infty}^{+\infty}\rmd x\, \frac{2\gamma_4\abs{\alpha_1}^2\sum_{j=5}^d\abs{b_j+s_j(-\rmi
x)}^2}{\pi\left(\gamma_4^{\;2}  +4\left(\mu-x\right)^2\right) \abs{\varkappa_0 +\rmi
x}^2}\,;
\end{equation}
here $\nu_4=  \nu$ and $\gamma_4=\kappa$ for heterodyning or $\nu_4= \nu_3$ and $\gamma_4=\varepsilon$ for homodyning.

Let us stress that  $S_{11}(\mu)+ S_{12}(\mu)$ is the contribution to the fluorescent light of the laser
wave $f(t)$, while $S_{2}(\nu_4+\mu)+S_{2}(\nu_4-\mu)$ is the contribution of the environment and of the
incoherent light. Moreover, by comparing Eqs.\ \eqref{Lambdaj} and \eqref{eq:S2}, we obtain
\begin{equation}\label{hd_Lambdaj}
\frac 1 {4\pi} \int_{-\infty}^{+\infty}S_{2}(\mu)\,\rmd \mu
=\frac{\abs{\alpha_1}^2}2\sum_{j =5}^d\Lambda_j.
\end{equation}

\subsubsection{Heterodyne spectrum}

In the case of heterodyning, the processes $f$ \eqref{eq:stim.ph} and $h$ \eqref{locosc} are independent and
this simplifies the computations of the fourth order moments involved in the expressions \eqref{S11} and
\eqref{S12}. In Appendix \ref{app:Hd} we show that
\begin{equation}\label{S=0}
S_{12}(\mu)=0,
\end{equation}
\begin{equation}\label{S1pm}
S_{11}(\mu)= S_{1}(\nu+\mu)+S_{1}(\nu-\mu),
\end{equation}
\begin{multline}\label{eq:h21}
S_{1}(\nu)= \abs{\alpha_1}^2\abs{\alpha_2}^2 \abs{g}^2\Biggl\{ \frac 1{\frac
{(\kappa+\varepsilon)^2} 4 + \left(\nu_3-\nu\right)^2}\\ {} \times\left[ \frac \kappa{\frac
{(\gamma_0+\varepsilon)^2} 4 + \left(\Delta \nu\right)^2}+ \frac
\varepsilon{\frac {(\gamma_0+\kappa)^2} 4 + \left(\nu-\nu_0\right)^2}\right]
\\ {}+
\frac {\kappa\varepsilon}{\left(\frac{(\gamma_0+\varepsilon)^2} 4 +
\left(\Delta \nu\right)^2\right)\left(\frac {(\gamma_0+\kappa)^2} 4 +
\left(\nu-\nu_0\right)^2 \right)}\\ {} \times\left[ \frac 1 {\gamma_0} +
\frac{\gamma_0+\kappa+\varepsilon}{\frac {(\kappa+\varepsilon)^2} 4 +
\left(\nu_3-\nu\right)^2}\right]\Biggr\}.
\end{multline}
Recall that $\Delta \nu= \nu_0-\nu_3$ is the detuning.

By explicit computations, we find also that the mean current \eqref{moc} does not contribute to the spectrum:
\begin{equation*}
\lim_{T\to +\infty} \frac 1 T\abs{\int_0^T
\rme^{\rmi \mu t}\Ebb_{\Pbb_{\phi_0}^T}[I(t)]\rmd t}^2=0.
\end{equation*}

\paragraph{Intensity.} Note that from Eqs.\ \eqref{Lambdaf} and \eqref{eq:h21} we get
\begin{equation}\label{hd_Lambdaf}
\frac 1 {4\pi} \int_{-\infty}^{+\infty}S_{1}(\mu)\,\rmd \mu
=\frac{\abs{\alpha_1}^2}2\,\Lambda_f.
\end{equation}
Equations \eqref{hd_Lambdaj} and \eqref{hd_Lambdaf} connect the intensities in direct detection and heterodyning,
a general relation already encountered in other systems such as two-level atoms \cite{BarL00,BarG09}.

\paragraph{Perfect local oscillator.} To summarize, the heterodyne spectrum is given by Eqs.\ \eqref{Gm1},
\eqref{Gm3}, \eqref{eq:S2}, \eqref{S=0}, \eqref{S1pm}, and \eqref{eq:h21} with $\nu_4=  \nu$ and $\gamma_4=\kappa$.
These formulas are somewhat involved, as they contain contributions from various sources of noise:
$\gamma_0$ is the natural with of the cavity mode, $\varepsilon$ is the width of the stimulating laser, and
$\kappa$ is the width of the local oscillator. The situation becomes more transparent in the limit of a perfect
local oscillator: $\kappa\downarrow 0$. By recalling that $
\frac \kappa {2\pi\left[\left(x+\nu\right)^2+\kappa^2/4\right]} \to \delta(x+\nu)$, in this limit
the heterodyne spectrum reduces to
\begin{subequations}
\begin{equation}
S_m(\mu)=\sum_{i=1}^2 \left[S_i(\nu+\mu) + S_i(\nu-\mu)\right],
\end{equation}
\begin{equation}
S_{1}(\nu\pm\mu)=\frac {\abs{\alpha_1}^2\abs{\alpha_2}^2}
{\left(\nu\pm\mu-\nu_0\right)^2+\gamma_0^{\;2}/4}\, S_f(\nu\pm\mu),
\end{equation}
\begin{equation}\label{h11_k=0}
S_{2}(\nu\pm\mu)=\frac {\abs{\alpha_1}^2} {\left(\nu\pm\mu-\nu_0\right)^2+
\gamma_0^{\;2}/4}\,S_Y(\nu\pm\mu);
\end{equation}
\end{subequations}
the spectra of the stimulating laser wave $f$ and of the incoherent noise $Y(t)$ are given by Eqs.\
\eqref{fspectrum2} and \eqref{Yspectrum2}. Let us recall that $\nu_0$ is the resonance frequency of our quantum system
and $\nu$ is the frequency of the local oscillator in the detection apparatus, which can be adjusted by the experimenter.
We can say that from the heterodyne output we can see part of the
input spectra  $S_f$ and $S_Y$ through a window of width $\gamma_0$ centered on $\nu_0$. In principle, in the case of
a bad cavity with a big width $\gamma_0$,  we can read the Markovian and non-Markovian character of the
various contributions to the dynamics from the heterodyne spectrum.

\subsubsection{Homodyne spectrum}
The homodyne spectrum is given again by the general formulas \eqref{def:sp}, \eqref{Gm1}, and \eqref{Gm2}.
Now, in the homodyne scheme, the local oscillator and the stimulating laser wave come from the same source,
in order to maintain phase coherence during the detection process and to get a phase sensitive measurement
procedure. However, this does not give coherence between the local oscillator and the various sources of noise
described by the process $Y(t)$ (thermal reservoirs, incoherent light, etc.) and, indeed, the related
contribution $S_2(\mu)$  to the homodyne spectrum is the same as in heterodyning and it is given by
Eq.\ \eqref{eq:S2} with the substitutions $\nu_4 \to \nu_3$ and $\gamma_4\to \varepsilon$.

The situation is different in the case of the components $S_{11}(\mu)$ \eqref{S11} and $S_{12}(\mu)$
\eqref{S12}, which contain both the stimulating laser wave $f$ \eqref{eq:stim.ph} and the local
oscillator \eqref{lo_homo}. To simplify the computations we consider only two limiting cases.
The first case is when the optical paths of laser wave and local oscillator are much larger than the coherence
length of the source; this is done by taking a large time delay $\Delta t$ and this washes out any phase sensitivity.
The second case is when the two optical paths are perfectly balanced ($ \Delta t =0$) and the phase sensitivity is
maximal. The general case is in between these two extremes.

\paragraph{The case $ \Delta t \to \pm \infty$.} By analyzing the expressions \eqref{S11} and \eqref{S12}, one
can see that in this limit $f$ and $h$ become stochastically independent and $S_{11}$ and $S_{12}$ become
the same as in the heterodyne case with the substitutions $\nu\to \nu_3$ and $\kappa\to \varepsilon$, i.e.
$S_{12}(\mu)=0$, $S_{11}(\mu)= S_{1}(\nu_3+\mu)+S_{1}(\nu_3-\mu)$ and
\begin{multline}\label{t0>>}
S_{1}(\nu_3\pm \mu)= \frac {\varepsilon\abs{\alpha_1}^2\abs{\alpha_2}^2
\abs{g}^2}{\varepsilon^2+ \mu^2}\biggl[ \frac 1{\frac {(\gamma_0+\varepsilon)^2} 4 +
\left(\Delta\nu\right)^2}
\\ {}+ \frac 1{\frac {(\gamma_0+\varepsilon)^2} 4 +
\left(\Delta\nu\mp\mu\right)^2}\biggr]+\left[ \frac 1 {\gamma_0} +
\frac{\gamma_0+2\varepsilon}{\varepsilon^2 + \mu^2}\right]
\\ {}\times
\frac {\varepsilon^2\abs{\alpha_1}^2\abs{\alpha_2}^2
\abs{g}^2}{\left(\frac{(\gamma_0+\varepsilon)^2} 4 +
\left(\Delta\nu\right)^2\right)\left(\frac {(\gamma_0+\varepsilon)^2} 4 +
\left(\Delta\nu\mp\mu\right)^2 \right)}.
\end{multline}

\paragraph{The case $ \Delta t =0$.} As shown in Appendix
\ref{app:homo}, in this case we have
\begin{multline}\label{S_hom}
S_{11}(\mu)+S_{12}(\mu)= \frac{2\abs{\alpha_1}^2\abs{\alpha_2}^2 \abs{g}^2}
{\frac{\left(\gamma_0+\varepsilon\right)^2}4+\left(\Delta
\nu\right)^2}\biggl\{\left(\cos \zeta\right)^2
4\pi \delta(\mu)\\
{}+\frac\varepsilon \gamma_0\, \RE \biggl[\left(\frac 1 {\frac{\gamma_0+\varepsilon}
2-\rmi \left(\Delta\nu-\mu\right)}+\frac 1 {\frac{\gamma_0+\varepsilon}
2-\rmi \left(\Delta\nu+\mu\right)}\right)\\
{}\times\left(1-\frac{\gamma_0 \rme^{2\rmi
\zeta}}{\gamma_0+2\varepsilon -2\rmi \Delta\nu}\right)\biggr]\biggr\},
\end{multline}
\begin{equation}
\zeta:=\arg \left(\frac{\alpha_1\, \overline{\alpha_2}}{\frac {\gamma_0+\varepsilon}
2-\rmi\Delta \nu}\right)+\theta.
\end{equation}

Let us stress that $ \Delta t =0$ means that the two optical paths are perfectly balanced and
homodyning allows for perfect interference, which generates a $\delta$ spike at zero
frequency. By varying $\abs{ \Delta t }$ from zero to infinity, one goes from \eqref{S_hom}
to \eqref{t0>>} and $S_{12}(\mu)=0$.

From Eq.\ \eqref{S_hom} we see that, by adjusting the phase $\zeta$, we can change the relative intensity of the
$\delta$ spike and the regular part. To simplify, let us consider only two extreme cases. We take always a vanishing
detuning, which implies, in particular, $\zeta=\arg \left(\alpha_1\, \overline{\alpha_2}\right)+\theta$. First we define
\[
l(\mu):= \frac{16\abs{\alpha_1}^2 \abs{\alpha_2}^2 \abs{g}^2 }{\gamma_0\left[\left(\frac{\gamma_0+\varepsilon}2\right)^2+\mu^2
\right]}\,.
\]
Then, for $\Delta\nu=0$ and $\zeta=\pm \pi/2$, we have
\[
S_{11}(\mu)+S_{12}(\mu)= \frac \varepsilon {\gamma_0+2\varepsilon}\, l(\mu).
\]
In the other extreme case, $\Delta\nu=0$ and $\zeta=0$ or $\pi$, we have
\[
S_{11}(\mu)+S_{12}(\mu)= \frac {\varepsilon^2  l(\mu)} {\left(\gamma_0+\varepsilon\right)
\left(\gamma_0+2\varepsilon \right)} +\frac \pi 2 \, \gamma_0 l(0)\delta(\mu) .
\]
Let us stress that this phase sensitivity is the characterizing feature of homodyne detection.

\section{Conclusions}\label{sec:concl}
In this article we have presented a SSE in which the involved operators are allowed to be random and to depend on the past; moreover, the driving noises can be colored diffusive processes and general counting processes, not only white noise and Poisson processes. This modification introduces memory without violating the complete positivity of the dynamics for the reduced state (the \emph{a priori} state or mean state). In this way we have an unravelling of a completely positive dynamics with memory. The main difference with respect the Markovian case is that now a random Liouville operator appears and this precludes having a simple closed equation for the mean dynamics.

By constructing positive operator-valued measures and instruments, we have also shown that our proposal is compatible with an interpretation in terms of continuous monitoring of the system; the axiomatic structure of quantum mechanics is respected. The key point in the construction is that the starting point is the \emph{linear SSE} and that its structure is such that the square norm of the solution is a \emph{martingale}.

On physical grounds, this theory allows various memory effects to be introduced in a consistent way. The most important one is the possibility of introducing measurement-based feedback, with delay, a fact that opens the way to a general treatment of quantum closed-loop control. Memory and feedback can be introduced also in the detection part (say in the local oscillator of a hetero- or homodyne detector) and this allows for a consistent treatment of adaptive measurements.

In order to understand the possibilities of the theory, we have introduced the simplest quantum system, a harmonic oscillator with a dynamics leaving invariant the coherent states. The possibility of using random terms in the Liouville operator allows the introduction various memory effects, such as
colored environments, producing non-white thermal-like effects with any spectral density. Moreover,
the stimulating light can be a laser with imperfections (we have taken the phase diffusion model of a laser wave) and/or incoherent light (thermal light, for instance).

As possible continuous monitoring, we first studied direct detection and showed how the detected intensity and the Mandel $Q$ parameter depend on the characteristics of the input (light and thermal effects).

Then we studied the spectrum of the hetero- or homodyne current. Here, the imperfections in the local oscillator can also be consistently introduced in the theory. The two spectra are affected by all the noises and characteristics of the dynamics and have somewhat complicated expressions; however, two features appear. The heterodyne spectrum reproduces a part of the spectra of the thermal noises and of the stimulating light and, so, in principle, it allows us to see directly the sources of non-Markovian effects. The homodyne spectrum turns out to be phase sensitive, as is known; by controlling the initial phase it is possible to change the field quadrature that is detected. The possibility of introducing a coherence length in the stimulating light and in the local oscillator allows a study of the influence of imperfections on homodyning; one sees that, as soon the apparatus is not perfectly balanced, the homodyne spectrum becomes the same as the heterodyne spectrum.

\appendix

\section{Some autocorrelation functions}

\subsection{The stimulating laser}\label{app:sl}
Let us consider the process \eqref{eq:stim.ph} modeling the stimulating laser
light. We can say that $f$ is a log-normal process, i.e., the exponential of a
(complex) Gaussian process. In stochastic calculus it is well known that
$\exp\left\{\rmi\sqrt{\varepsilon}\, B_{3}(t)+\varepsilon  t/2\right\}$ is a mean-1 martingale, which gives the mean function
\begin{equation}\label{mom:f}
\Ebb_{\Qbb}\left[f(t)\right] =g\rme^{-\rmi \nu_3 t-\varepsilon t/2}
\end{equation}
and the autocorrelation functions
\begin{equation}\label{mom:ffbar}
\Ebb_{\Qbb}\left[f(r)\,\overline{f(s)}\right] = \abs{g}^2\rme^{\rmi
\nu_3(s-r)-\varepsilon \abs{s-r}/2},
\end{equation}
\begin{equation}\label{mom:ff}
\Ebb_{\Qbb}\left[f(r)f(s)\right] = g^2\rme^{-\left(\rmi \nu_3+\varepsilon
/2\right)(s+r)-\varepsilon (s\wedge r)},
\end{equation}
where $a\wedge b$ is the minimum between $a$ and $b$. In the proof of Eqs.\
\eqref{mom:ffbar} and \eqref{mom:ff} one has to use \eqref{mom:f} and the
independence of the increments of the Wiener process.

Let us stress that from \eqref{mom:ffbar} we get immediately the spectrum
\eqref{fspectrum2}, while from \eqref{mom:f} and \eqref{mom:ff} we get
\[
\lim_{T\to +\infty}
\frac 1 T \abs{\int_0^T \rme^{\rmi \mu t} \,\Ebb_\Qbb \left[f(t)\right]\rmd t}^2=0,
\]
\[
\lim_{T\to +\infty}
\frac 1 T \,\Ebb_\Qbb \left[\left(\int_0^T \rme^{\rmi \mu t} f(t)\,\rmd t\right)^2\right]=0.
\]

We shall need also the autocorrelation function of the related process $U_f(t)$ for large times.
From \eqref{Uf}, \eqref{mom:ffbar}, and \eqref{mom:ff} we obtain
\begin{equation}\label{UUaut}
\lim_{t_0\to+\infty}\Ebb_\Qbb\left[ U_f(t+t_0) U_f(s+t_0)\right]=0,
\end{equation}
\begin{multline}\label{UUbaut}
\lim_{t_0\to+\infty}\Ebb_\Qbb\left[ U_f(t+t_0) \overline{U_f(s+t_0)}\right] \\
{}=\frac{ \abs{\alpha_2}^2 \abs{g}^2}{\gamma_0} \biggl\{ \frac{\rme^{-\overline{\varkappa_0}
\left(t-s\right)}}{\overline{\varkappa_0}+\varkappa_3}
+\frac{\rme^{-\overline{\varkappa_3}\left(t-s\right)}}{\varkappa_0+\overline{\varkappa_3}} \\
{}+ \frac{\rme^{-\overline{\varkappa_3}\left(t-s\right)}-
\rme^{-\overline{\varkappa_0}\left(t-s\right)}}{\overline{\varkappa_0}-\overline{\varkappa_3}} \biggr\}, \qquad t\geq s,
\end{multline}
where $\varkappa_3=\frac\varepsilon 2-\rmi \nu_3$.

\subsection{Autocorrelation functions of the process $Y$}\label{A4}
Let us consider the Gaussian process $Y$
defined by Eqs.\ \eqref{Y(t)} and \eqref{Xj(t)}; we can write
\begin{multline*}
\int_0^T \rme^{\rmi \mu t}\rmd Y(t)= \sum_{j=5}^d\int_0^T \rmd B_j(t) \, \rme^{\rmi \mu t}
\\ {}\times\left(b_j+ \int_0^{T-t}\rmd s \, \rme^{\rmi \mu s }c_j(s)\right).
\end{multline*}
Then, by using the It\^o isometry we get
\begin{multline*}
\frac 1 T \,\Ebb_\Qbb \left[\abs{\int_0^T \rme^{\rmi \mu t} \rmd Y(t)}^2\right]
\\ {}=\frac 1 T\sum_{j=5}^d\int_0^T\rmd t \abs{b_j+\int_0^t \rmd s \, c_j(s)\rme^{ \rmi
\mu s}}^2.
\end{multline*}
From the limit $T\to +\infty$ we get immediately the Laplace transform \eqref{LapTc}
inside the square modulus and \eqref{Yspectrum2} is proved.

The expression \eqref{Y(t)} of the process $Y$ contains the function $c_j(t)$ and
its spectrum involves the Laplace transform $s_j(z)$ \eqref{LapTc} of $c_j$. In the
following, we shall need also the Fourier inversion formula:
\begin{equation}\label{invLapl}
c_j(t)=\frac 1{2\pi}\int_{-\infty}^{+\infty}\rme^{-\rmi \mu  t} s_j(-\rmi \mu)\rmd \mu, \qquad t>0.
\end{equation}

Similarly to  \eqref{LapTc} and \eqref{invLapl}, for the function  $g_j(t)$ \eqref{xi2} we have
\begin{equation}\label{gtoz}
\int_0^{+\infty} \rme^{-z t} g_j(t)\,\rmd t
= \frac{{b_j}+ {s_j(z)}}{\overline{\varkappa_0}+z}, \qquad \RE z\geq 0,
\end{equation}
\begin{equation}\label{ztog}
g_j(t)= \frac 1 {2 \pi} \int_{-\infty}^{+\infty} \rme^{-\rmi x t }\, \frac{{b_j}+
{s_j(-\rmi x)}}{\overline{\varkappa_0}-\rmi x}\, \rmd x , \quad t>0.
\end{equation}

For the related process $U_Y(t)$, from Eqs.\ \eqref{UY}, \eqref{gtoz}, and \eqref{ztog} we obtain,  with $t\geq s$,
\begin{multline}\label{YUUbaut}
\lim_{t_0\to+\infty}\Ebb_\Qbb\left[ U_Y(t+t_0) \overline{U_Y(s+t_0)}\right]\\
{} =\frac{1}{2\pi} \int_{-\infty}^{+\infty}  \rme^{-\rmi x\left(t-s\right)}\sum_{j=5}^d
\frac{\abs{b_j+s_j(-\rmi x)}^2} {\abs{ \varkappa_0 + \rmi x}^2}\, \rmd x,
\end{multline}
\begin{multline}\label{YUUaut}
\lim_{t_0\to+\infty}\Ebb_\Qbb\left[ U_Y(t+t_0) U_Y(s+t_0)\right] =\frac{1}{2\pi} \int_{-\infty}^{+\infty}
\rmd x\\ {}\times \rme^{-\rmi x\left(t-s\right)}\sum_{j=5}^d  \frac{\left[b_j+s_j(-\rmi x)\right]
\left[b_j+s_j(\rmi x)\right]} { {\overline{\varkappa_0}}^2 +  x^2}\,.
\end{multline}

\subsection{The local oscillator in the heterodyne scheme}
For the moments of the local oscillator in the heterodyne measurement scheme
\eqref{locosc} we have formulas analogous to those of Appendix \ref{app:sl}:
\begin{equation}\label{mom:h}
\Ebb_{\Qbb}\left[h(t)\right] =
\rme^{\rmi \vartheta}\rme^{-\rmi
\nu t-\kappa  t/2},
\end{equation}
\begin{equation}\label{mom:hhbar}
\Ebb_\Qbb\left[\overline{h(t)}
\,h(s)\right]=\rme^{\rmi\nu(t-s)-\kappa \abs{t-s}/2},
\end{equation}
\begin{equation}\label{mom:hbhb}
\Ebb_\Qbb\left[\overline{h(t)}
\,\overline{h(s)}\right]=\rme^{-2\rmi\vartheta}\rme^{
\left(\rmi\nu-\kappa /2 \right)\left(t+s\right)-\kappa(t\wedge s)}.
\end{equation}

\subsection{The local oscillator in the homodyne scheme}
In the homodyne measurement scheme the local oscillator $h$ is proportional to the
shifted stimulating laser $f$ and it is given by Eq.\ \eqref{lo_homo}, i.e.,
\begin{equation}\label{hhom}
h(t)=\frac{g\rme^{\rmi \theta}}{\abs{g}}\, \rme^{-\rmi \nu_3 \left(t-\Delta t\right) +
\rmi\sqrt{\varepsilon}\, B_{3}(t- \Delta t )}.
 \end{equation}
As before we get
\begin{equation}\label{mom:h'}
\Ebb_{\Qbb}\left[h(t)\right] = \frac{g\rme^{\rmi \theta}}{\abs{g}}\,
\rme^{-\rmi
\nu_3\left( t- \Delta t \right)-\varepsilon \abs{ t- \Delta t }/2},
\end{equation}
\begin{equation}\label{mom:hhbar'}
\Ebb_\Qbb\left[\overline{h(t)}
\,h(s)\right]=\rme^{\rmi\nu_3(t-s)-\varepsilon \abs{t-s}/2},
\end{equation}
\begin{equation}\label{mom:hbhb'}
\Ebb_\Qbb\left[\overline{h(t)} \,\overline{h(s)}\right]=\left(\frac{g\rme^{-\rmi
\theta}}{\abs{g}}\right)^2 \rme^{ \left(\rmi\nu_3-\frac\varepsilon 2
\right)\left(t+s-2 \Delta t \right)-\varepsilon\left(t\wedge s - \Delta t \right)}.
\end{equation}

In the homodyne case we need also the two fourth order moments
\begin{widetext}
\begin{equation}\label{ffbhbh}
\Ebb_\Qbb\left[f(r_1)\overline{f(r_2)}\,\overline{h(t)}
\,h(s)\right]=\abs{g}^2\exp\left\{\rmi\nu_3 \left(t-r_1-s+r_2\right)
-\frac \varepsilon 2\,\delta(r_1,t- \Delta t ,s- \Delta t ,r_2)\right\},
\end{equation}
\begin{equation}\label{ffhbhb}
\Ebb_\Qbb\left[f(r_1)f(r_2)\overline{h(t)}
\,\overline{h(s)}\right]=\abs{g}^2
\exp\left\{\rmi\nu_3 \left(t-r_1+s-r_2-2 \Delta t \right)
-\frac \varepsilon 2\,\delta(r_1,t- \Delta t ,r_2,s- \Delta t )\right\},
\end{equation}
where
\[
\exp\left\{
-\frac \varepsilon 2\,\delta(t_1,t_2,t_3,t_4)\right\}=\Ebb_\Qbb\left[\exp\left\{
\rmi\sqrt{ \varepsilon}\bigl(B_3(t_1)-B_3(t_2)+B_3(t_3)-B_3(t_4)\bigr)\right\}\right].
\]
By some long, but straightforward computations we get
\[
\delta(t_1,t_2,t_3,t_4)= \begin{cases}
\abs{t_4-t_3}+\abs{t_2-t_1}, & \text{for } t_1\vee t_2 < t_3\wedge t_4 \text{ or }
t_3\vee t_4 < t_1\wedge t_2,
\\
\abs{t_4-t_1}+\abs{t_3-t_2}, & \text{for } t_1\vee t_4 < t_2\wedge t_3 \text{ or }
t_2\vee t_3 < t_1\wedge t_4,
\\
\abs{t_4-t_1}+3\abs{t_3-t_2}, & \text{for } t_4<t_2 < t_3< t_1 \text{ or }
t_1<t_3 < t_2< t_4,
\\
\abs{t_4-t_3}+3\abs{t_2-t_1}, & \text{for } t_4<t_2 < t_1< t_3 \text{ or }
t_3<t_1 < t_2< t_4,
\\
\abs{t_2-t_3}+3\abs{t_4-t_1}, & \text{for } t_2<t_4 < t_1< t_3 \text{ or }
t_3<t_1 < t_4< t_2,
\\
\abs{t_2-t_1}+3\abs{t_3-t_4}, & \text{for } t_2<t_4 < t_3< t_1 \text{ or }
t_1<t_3 < t_4< t_2.
\end{cases}
\]
\end{widetext}

\section{Detection}
\subsection{Direct detection}
\subsubsection{The mean counting intensity}\label{mci}
The expression of $\xi(t)$ is given by \eqref{xi1}; moreover, the processes $U_f$ \eqref{Uf} and $U_Y$
\eqref{UY} are independent and $U_Y$ is Gaussian with mean zero. So we have
\[
\Ebb_\Qbb[\abs{\xi(t)}^2]=\Ebb_\Qbb[\abs{U_f(t)}^2]+\Ebb_\Qbb[\abs{U_Y(t)}^2].
\]
From \eqref{UUbaut} and \eqref{YUUbaut} we get
\begin{multline}\label{Mlim1}
\Ebb_\Qbb[\abs{\xi(t)}^2]=
\frac{\abs{\alpha_2}^2 \abs{g}^2 \left(\gamma_0+\varepsilon  \right)}{\gamma_0\left( \frac{\left(\gamma_0
+\varepsilon\right)^2}4+\left(\Delta\nu\right)^2\right)}\\ {}
+\frac
{1}{2\pi}\sum_{j=5}^d \int_{-\infty}^{+\infty}  \frac
{\abs{b_j+s_j(-\rmi x)}^2}{\abs{\varkappa_0+\rmi x}^2} \,\rmd x .
\end{multline}
From Eq.\ \eqref{Mlim1} we obtain the expression for $\Lambda$ \eqref{Lambda}.

\subsubsection{The Mandel parameter}\label{app:Qpar}
By using Eqs.\ \eqref{Qt0t2}, \eqref{Qt}, \eqref{xi1}, and \eqref{UUaut} and by recalling that the process $U_Y$ is
Gaussian with zero mean and $U_Y(s)$ and $U_f(r)$ are independent, we get
\begin{multline*}
Q(t)=\frac{2\lambda\abs{\beta}^2\gamma_0}{\Lambda
t} \\ {} \times
\lim_{t_0\to + \infty}\int_{t_0}^{t+t_0}\rmd s\int_{t_0}^{s}\rmd
r \biggl\{\abs{
\Ebb_\Qbb\left[\overline{U_Y(s)} \, U_Y(r) \right]}^2
\\{}+ \abs{ \Ebb_\Qbb\left[U_Y(s)
\, U_Y(r) \right]}^2 +
\Cov_\Qbb\left[
\abs{U_f(s)}^2,\abs{U_f(r)}^2\right]
\\ {}+2\RE
\Ebb_\Qbb\left[ U_f(s) \,\overline{U_f(r)}\right] \Ebb_\Qbb\left[\overline{U_Y(s)}
\, U_Y(r) \right]\biggr\}.
\end{multline*}
We see that $Q(0)=0$; moreover, for large times the integrand depends only on the difference $s-r$ and one
can check that $\dot Q(t)\geq 0$. So $Q(t)$ is non decreasing and
\begin{multline}\label{Qtgen}
Q(t)=\frac{2\lambda\abs{\beta}^2\gamma_0}{\Lambda
} \int_{0}^{t}\rmd s \left(1-\frac st\right)
\\ {} \times
\lim_{t_0\to + \infty}\biggl\{\abs{
\Ebb_\Qbb\left[\overline{U_Y(t_0+s)} \, U_Y(t_0) \right]}^2
\\ {}
+ \abs{ \Ebb_\Qbb\left[U_Y(t_0+s)
\, U_Y(t_0) \right]}^2\\ {} +2\RE
\Ebb_\Qbb\left[ U_f(t_0+s) \,\overline{U_f(t_0)}\right] \Ebb_\Qbb\left[\overline{U_Y(t_0+s)}
\, U_Y(t_0) \right]
\\ {}
+
\Cov_\Qbb\left[
\abs{U_f(t_0+s)}^2,\abs{U_f(t_0)}^2\right]
\biggr\}.
\end{multline}
By using \eqref{UUbaut}, \eqref{YUUbaut}, and \eqref{YUUaut} and the analogs of \eqref{ffbhbh} and \eqref{ffhbhb},
one sees that $\lim_{t\to + \infty}Q(t)$ exists and it would be possible to compute $Q(t)$, but its general
expression is very involved and not particularly instructive.

\subsection{Homo- and heterodyne spectra}\label{app:sp}

By \eqref{hhI(t)} and the independence of $W_1$ and $m_1$ under the physical probability $\Pbb_{\phi_0}^T$,
we get, by some changes of integration variables,
\begin{multline*}
\frac 1 T\, \Ebb_{\Pbb_{\phi_0}^T}\left[\abs{\int_0^T
\rme^{\rmi \mu t}I(t)\rmd t}^2\right]\\ {}= \frac 1 T\int_0^T\rmd s \abs{\int_0^s \rmd t \,
\rme^{\rmi \mu t}F(t)}^2
\\ {}
+
\frac 1 T \,\Ebb\left[ \abs{\int_0^T \rmd t \,\rme^{\rmi \mu t}m_1(t) \int_0^{T-t} \rmd s \,
\rme^{\rmi \mu s}F(s)}^2\right].
\end{multline*}
Then, in the limit $T\to +\infty$, we obtain Eqs.\ \eqref{Gm1} and \eqref{Gm2}.

To compute the spectrum $S_m(\mu)$ \eqref{Gm2} we have to take the expression \eqref{m1} of $m_1$ and to use
Eqs.\ \eqref{xi1}-\eqref{xi2}. By recalling that
$U_Y$ is Gaussian, with zero mean and independent of $h$ and $f$, by some computations we arrive at the
expression $S_m(\mu)= S_{11}(\mu)+S_{12}(\mu) + \tilde S_{21}(\mu) +\tilde S_{22}(\mu)$ with
\begin{widetext}
\begin{multline}\label{S11}
S_{11}(\mu) =\lim_{T\to +\infty}\frac {4\abs{\alpha_1}^2\abs{\alpha_2}^2} T
\,\RE\int_0^T\rmd t\int_0^t \rmd s \, \cos\left( \mu \left(t-s\right)\right)\rme^{-\overline{\varkappa_0}\,
t-\varkappa_0 s}\biggl\{\int_0^s \rmd r_1
\int_0^{r_1} \rmd r_2\,\rme^{\overline{\varkappa_0}r_2+\varkappa_0r_1}
\\
{}\times\Ebb_\Qbb\left[\overline{f(r_1)}\,f(r_2)\,\overline{h(t)} \,h(s)\right]+\left(\int_s^t \rmd r_1 \int_0^s
\rmd r_2+\int_0^s \rmd r_1
\int_0^{r_1} \rmd r_2\right)
\rme^{\overline{\varkappa_0}r_1 +\varkappa_0r_2}
\Ebb_\Qbb\left[f(r_1)\overline{f(r_2)}\,\overline{h(t)} \,h(s)\right]
\biggr\},
\end{multline}
\begin{multline}\label{S12}
S_{12}(\mu)=\lim_{T\to +\infty}\RE\, \frac {4 \overline{\alpha_1}^2{\alpha_2}^2} T
\int_0^T\rmd t\int_0^t \rmd s \, \cos\left(\mu \left(t-s\right)\right)
\biggl(\int_s^t \rmd r_1 \int_0^s \rmd r_2\\ {}+2\int_0^s \rmd r_1 \int_0^{r_1} \rmd
r_2 \biggr)\rme^{-\overline{\varkappa_0}(t+s-r_1-r_2)}
\Ebb_\Qbb\left[f(r_1)f(r_2)\overline{h(t)} \,\overline{h(s)}\right],
\end{multline}
\begin{equation}\label{S21}
\tilde S_{21}(\mu)= \lim_{T\to +\infty}\frac {4\abs{\alpha_1}^2} T\sum_{j=5}^d\RE
\int_0^T\rmd t\int_0^t \rmd s \int_0^s \rmd r\, \cos\left( \mu
\left(t-s\right)\right) g_j(t-r)\overline{g_j(s-r)}\,\Ebb_\Qbb\left[\overline{h(t)}
\,h(s)\right],
\end{equation}
\begin{equation}\label{S22}
\tilde S_{22}(\mu)= \lim_{T\to +\infty}\RE  \frac {4\overline{\alpha_1}^2} T
\sum_{j=5}^d\int_0^T\rmd t\int_0^t \rmd s \int_0^s \rmd r \,\cos\left(\mu
\left(t-s\right)\right) g_j(t-r) g_j(s-r)\Ebb_\Qbb\left[\overline{h(t)}
\,\overline{h(s)}\right].
\end{equation}

Let us consider now both heterodyning and homodyning; the moments \eqref{mom:hhbar} and \eqref{mom:hhbar'}
can be written in a unified way as
\[
\Ebb_\Qbb \left[ \overline{h(t)}\, h(s)\right]= \rme^{-\varkappa_4\left(t-s\right)}, \qquad t\geq s ,
\]
with $\varkappa_4= \frac \kappa 2 -\rmi \nu$ for heterodyning, $\varkappa_4= \frac \varepsilon 2 -\rmi \nu_3$
for homodyning. Then, from \eqref{S21} we obtain
\begin{equation}\label{S_21}
\tilde S_{21}(\mu)= 4\abs{\alpha_1}^2 \sum_{j=5}^d \RE \int_0^{+\infty}\rmd t \int_0^t\rmd
s\, \rme^{-\varkappa_4\left(t-s\right)}
\cos\big(\mu(t-s)\big)g_j(t)\overline{g_j(s)}.
\end{equation}
By inserting \eqref{ztog} into \eqref{S_21} and, then, by using \eqref{gtoz} for $\overline{g_j(s)}$, we get
$\tilde S_{21}(\mu)=S_{2}(\nu_4+\mu)+S_{2}(\nu_4-\mu)$ with $S_{2}(\mu)$ given by \eqref{eq:S2}.

By similar computations, from \eqref{S22} we find that $\tilde S_{22}(\mu)$ vanishes for both the choices \eqref{lo_homo} and \eqref{locosc} for the local oscillator $h(t)$; this ends the proof of \eqref{Gm3} and \eqref{eq:S2}.

\subsubsection{Heterodyne detection}\label{app:Hd}
By using \eqref{mom:ffbar} and \eqref{mom:hhbar} inside \eqref{S11}, we get \eqref{S1pm} with
\begin{multline}\label{S1a}
S_{1}(\nu) =\lim_{T\to + \infty}  \frac{2\abs{\alpha_1}^2\abs{\alpha_2}^2
\abs{g}^2}{T}\RE \int_0^{T}\rmd t \int_0^{t}\rmd s\, \rme^{\left(\rmi \nu -\frac
\kappa 2\right)(t-s)}\biggl\{\biggl( \int_s^t\rmd r_1\int_0^s\rmd r_2 + \int_0^s\rmd
r_1\int_0^{r_1}\rmd r_2\biggr)
\\ {}\times
\rme^{-\overline{\varkappa_0}\left(t-r_1\right) -\varkappa_0 \left(s-r_2\right)
+\rmi \nu_3\left(r_2-r_1\right)- \frac \varepsilon 2\left(r_1-r_2\right)} +
\int_0^s\rmd r_1\int_0^{r_1}\rmd r_2\,
\rme^{-\overline{\varkappa_0}\left(t-r_2\right) -\varkappa_0 \left(s-r_1\right)
+\rmi \nu_3\left(r_1-r_2\right)- \frac \varepsilon 2\left(r_1-r_2\right)}\biggr\}
\\ {}= \RE\frac{2\abs{\alpha_1}^2\abs{\alpha_2}^2 \abs{g}^2}{\overline{\varkappa_0} +\frac \kappa 2 -\rmi \nu}
\biggl\{ \frac 1
{\varkappa_0
+\frac \varepsilon 2 +\rmi \nu_3  }\left[\frac 1 {\frac {\kappa+\varepsilon}
2 +\rmi \left(\nu_3-\nu\right) }+\frac 1{\gamma_0}\right]
+ \frac 1
{\left(\overline{\varkappa_0} +\frac \varepsilon 2 -\rmi \nu_3\right)
\gamma_0 }\biggr\}.
\end{multline}
By explicit computations, from \eqref{S1a} we get \eqref{eq:h21}.
Similarly, by \eqref{S12}, \eqref{mom:ff}, and \eqref{mom:hbhb} we obtain $S_{12}(\mu)=0$.

\subsubsection{Homodyning, $ \Delta t =0$} \label{app:homo}
By inserting the expression \eqref{ffbhbh} into \eqref{S11} and \eqref{ffhbhb} into \eqref{S12}, we obtain
\begin{equation}\label{t00}
S_{11}(\mu)= \frac{\abs{\alpha_1}^2\abs{\alpha_2}^2
\abs{g}^2}{\frac{\left(\gamma_0+\varepsilon\right)^2}4+\left(\Delta \nu\right)^2}
\biggl\{4\pi
\delta(\mu)+\frac{\varepsilon(\gamma_0+\varepsilon)}{\gamma_0}\biggl[\frac 1{\frac{\left(\gamma_0+
\varepsilon\right)^2}4+\left(\mu+
\Delta\nu\right)^2} +\frac 1{\frac{\left(
\gamma_0 +\varepsilon\right)^2}4+\left(\mu- \Delta\nu\right)^2}\biggr]\biggr\},
\end{equation}
\begin{equation}\label{t11}
S_{12}(\mu)= \RE \,\frac{\abs{g}^2 \rme^{2\rmi
\theta}{\alpha_1}^2{\overline{\alpha_2}}^2 }{\left(\varkappa_0+\rmi \nu_3
+\frac\varepsilon 2\right)^2}\biggl\{4\pi \delta(\mu) - \frac
{\varepsilon}{\varkappa_0+\rmi \nu_3 +\varepsilon}
\biggl[\frac 1
{\varkappa_0+\rmi \left(\nu_3 +\mu\right)+\frac\varepsilon 2}+\frac 1
{\varkappa_0+\rmi \left(\nu_3 -\mu\right)+\frac \varepsilon 2}\biggr]\biggr\}.
\end{equation}
\end{widetext}
In doing the computations we have used the well known approximations of the Dirac $\delta$ $ \frac{\sin \mu T}
{\pi \mu}\to \delta(\mu)$ and $
\frac{2\left(\sin \mu T/2\right)^2}{\pi T \mu^2}=\frac{1-\cos \mu T}{\pi T \mu^2}
\to \delta(\mu)$. By adding the expressions \eqref{t00} and \eqref{t11} we get \eqref{S_hom}.

\end{document}